\documentstyle[12pt,aaspp4]{article}

\begin{document}

\title{Diverse Supernova Sources for the  $r$-Process}
\author{Y.-Z. Qian and P. Vogel}
\affil{Physics Department, California Institute of Technology,
       Pasadena, CA 91125}
\authoremail{yzqian@citnp.caltech.edu, vogel@lamppost.caltech.edu}
\and
\author{G. J. Wasserburg}
\affil{The Lunatic Asylum,
Division of Geological and Planetary Sciences, California
       Institute of Technology, Pasadena, CA 91125}

\begin{abstract}
We present a simplified analysis using equations for the charge
flow, which include $\nu_e$ capture, for the production of $r$-process
nuclei in the context of the recent supernova hot bubble model.
The role of $\nu_e$ capture in speeding up the charge
flow, particularly at the closed neutron shells, is studied together 
with the $\beta$-flow at
freeze-out and the effect of neutrino-induced neutron emission
on the abundance pattern after freeze-out. It is shown
that a semi-quantitative agreement with the gross solar $r$-process 
abundance pattern from the peak at mass number
$A\sim 130$, through the peak at $A\sim 195$, and up to the region 
of the actinides can be obtained by
a superposition of two distinctive kinds of $r$-process
events. These correspond to a low frequency case L and a high frequency
case H, which takes into account 
the low abundance of $^{129}$I and the high
abundance of $^{182}$Hf in the early solar nebula. The lifetime of
$^{182}$Hf ($\tau_{182}\approx 1.3\times 10^7$ yr) associates
the events in case H
with the most common Type II supernovae.
These events would be mainly
responsible for the $r$-process nuclei near and above
$A\sim 195$. They would also
make a significant amount of the nuclei between
$A\sim 130$ and 195, including $^{182}$Hf, but very little 
$^{129}$I. In order to match the solar $r$-process abundance pattern
and to satisfy the $^{129}$I and $^{182}$Hf constraints, 
the events in case L, which would make the $r$-process nuclei near
$A\sim 130$ and the bulk of those between $A\sim 130$ and 195,
must occur $\sim 10$ times
less frequently but eject $\sim 10$--20 times more $r$-process material
in each event.

Assuming that all of the supernovae 
producing $r$-process nuclei represent a similar overall process,
we speculate that the usual neutron
star remnants, and hence prolonged ejection of $r$-process material,
are associated with the events in case L. We further speculate that
the more frequently occurring events in case H have
ejection of other $r$-process material terminated by
black hole formation during the neutrino cooling phase of
the protoneutron star.
This suggests that there is now an inventory of $\sim 5\times 10^8$
black holes with masses $\sim 1\ M_\odot$ and $\sim 5\times 10^7$
neutron stars resulting from supernovae in the Galaxy. 
This $r$-process model would have little
effect on the estimates of the supernova contributions to the non-$r$-process
nuclei.
\end{abstract}

\section{Introduction}
Approximately half of the heavy elements with mass
number $A>70$ and all of the actinides in the solar system
are believed to be produced in the $r$-process.
The fundamental $r$-process theory of 
Burbidge et al.\markcite{bbfh} (1957)
and Cameron\markcite{cam1} (1957) successfully 
explains the gross features of the 
solar $r$-process abundance distribution,
such as the existence of abundance peaks at $A\sim 80$, 130, and
195. On the other hand, it remains to be established
where the $r$-process occurs and especially
how many different kinds of $r$-process events contributed to the
solar $r$-process abundances. Major advances have been made in
calculating $r$-process nucleosynthesis in supernovae
(see e.g., Woosley et al.\markcite{woosley2} 1994) and in
using a wide range of model parameters to obtain yields that
approximate the solar $r$-process abundances
(see e.g., Kratz et al.\markcite{kratz} 1993). There has been a
tendency to ascribe all the $r$-process nuclei to a single
kind of $r$-process events (but see 
Goriely \& Arnould\markcite{goriely} 1996). However, most
astrophysical models have difficulty in producing all the
$r$-process abundance peaks from a single source, and
the parametric studies certainly do not point to a single
kind of $r$-process events.

With the recent progress in both observation and theory, there is
a growing consensus that Type II supernovae 
are the most probable $r$-process
site. The detection of $r$-process elements in the extremely
metal-poor halo star CS 22892-052 by 
Sneden et al.\markcite{sneden} (1996) argues
that the $r$-process is primary, already operating
in the early history of the Galaxy. Studies of Galactic chemical
evolution (Mathews, Bazan, \& Cowan\markcite{mathews} 1992) 
show that the enrichment
of the $r$-process elements in the Galaxy is consistent with low
mass Type II supernovae being the $r$-process site. Furthermore, it
has been proposed that the $r$-process occurs in the neutrino-heated
ejecta from the hot protoneutron star produced in a Type II supernova
(Woosley \& Baron\markcite{woosley1} 1992;
Woosley \& Hoffman\markcite{woosley2} 1992; 
Meyer et al.\markcite{meyer1} 1992; 
Woosley et al.\markcite{woosley3} 1994). 
While this so-called ``hot bubble''
$r$-process model has
some deficiencies, especially the need for very high entropies
that might be hard to obtain 
(Witti, Janka, \& Takahashi\markcite{witti} 1994; 
Takahashi, Janka, \& Witti\markcite{taka} 1994; 
Qian \& Woosley\markcite{qian1} 1996;
Hoffman, Woosley, \& Qian\markcite{hoffman} 1997),
it also has several attractive features.
For example, the amount of ejecta from the hot bubble is consistent
with the expected amount of $r$-process material from each supernova
(Woosley et al.\markcite{woosley3} 1994), 
and unlike the entropy, can be 
understood quite well in terms of a simple neutrino-driven wind
model (Qian \& Woosley\markcite{qian1} 1996). 
In addition, it has
been shown that the intense neutrino flux in this kind of 
$r$-process model can have important effects on the nucleosynthesis
(Meyer\markcite{meyer2} 1995;
Fuller \& Meyer\markcite{fuller} 1995;
McLaughlin \& Fuller\markcite{mclau1} 1996;
Qian et al.\markcite{qian2} 1997;
Haxton et al.\markcite{haxton} 1997;
McLaughlin \& Fuller\markcite{mclau2} 1997).
In particular, the typical neutrino fluences through the ejecta
may lead to identifiable signatures in the $r$-process abundance
pattern, thus providing a way to reveal 
the conditions at the $r$-process site
(Qian et al.\markcite{qian2} 1997; 
Haxton et al.\markcite{haxton} 1997).

Regardless of the astrophysical site, two things are needed for
an $r$-process to work: the neutrons and the seed nuclei to
capture them. In fact, the potential of an astrophysical
environment to be the $r$-process site can be gauged by a
crucial quantity, the neutron-to-seed ratio.
If one always starts from more or less the same seed nuclei,
different neutron-to-seed ratios are required to produce
the entire solar $r$-process nuclear abundance
distribution. One can then ask
whether different $r$-process
nuclei are made in completely different astrophysical environments
(e.g., Type II supernovae vs. neutron star coalescence) or in
similar environments but just
with different neutron-to-seed ratios.
Because the $r$-process abundance distribution in CS 22892-052
agrees with that in the solar system quite well
(Sneden et al.\markcite{sneden} 1996), and the solar
$r$-process abundance distribution does not have 
sudden jumps as a function of 
the mass number $A$, it may be more natural
to expect that all $r$-process nuclei come from similar
environments (e.g., the hot bubble regions in Type II supernovae).
Hereafter, we
refer to the production of $r$-process nuclei
in a specific environment
with a certain distribution of neutron-to-seed ratios as
an $r$-process ``event.'' The simplest scenario would be
that all $r$-process nuclei are produced in
a unique kind of $r$-process events with a generic
abundance pattern.
In that case,
the solar $r$-process abundance distribution merely
reflects the distribution of neutron-to-seed ratios
characteristic of these unique $r$-process events.

However, 
Wasserburg, Busso, \& Gallino\markcite{wass} (1996) pointed out
that the above minimal approach to account for the solar
$r$-process abundance distribution is not consistent with
the meteoritic abundance ratios
$^{129}$I/$^{127}$I and 
$^{182}$Hf/$^{180}$Hf in the
early solar system. 
These authors showed that the 
$r$-process events contributing to $^{182}$Hf 
were fully consistent with
the uniform production of $^{232}$Th, $^{235}$U, $^{238}$U,
and $^{244}$Pu up until the time when the solar system was
formed. However, such a rather uniform production would
grossly overproduce $^{129}$I (by a factor of $\sim 50$)
and $^{107}$Pd (by a factor of $\sim 30$). 
Consequently, they argued that
there should be diverse sources for the $r$-process,
one of which produced the $r$-process nuclei above
$A\sim 140$ and another producing those at lower $A$
with a smaller frequency.

In order to 
account for the solar
$r$-process abundance distribution and to
accommodate the meteoritic data on
$^{129}$I and $^{182}$Hf at the same time,
we consider in this paper a minimal scenario where
two kinds of $r$-process events contribute to the solar
$r$-process abundances near and above $A\sim 130$. 
Using simplified treatment of the
$r$-process and taking into account other
constraints, 
we show that the main features of the solar $r$-process
abundance distribution from the peak at $A\sim 130$,
through the peak at $A\sim 195$, and up
to the region of the actinides
can be reproduced by a reasonable
superposition of these two kinds of $r$-process events.
In \S2, we describe the hot bubble $r$-process
model and our simplified
$r$-process calculation in the context of this model.
We also discuss the constraints on our $r$-process
calculations from
the observed solar $r$-process abundance distribution,
the meteoritic data on $^{129}$I and $^{182}$Hf, and
considerations of various neutrino effects.
In \S3, we present our results, and
in \S4 we discuss
their implications
for the nature and frequencies of 
the supernovae associated with these two kinds of
$r$-process events.

\section{Supernova $r$-Process Model}
In the following discussion we make the general
assumption that the $r$-process occurs in the
neutrino-heated supernova ejecta 
(or neutrino-driven winds) from the hot
protoneutron star, as in the hot bubble $r$-process
model. In this model, a neutron-rich mass element
expands due to the heating by neutrinos emitted
from the protoneutron star. The mass element initially
is composed of free nucleons. As it moves away from
the protoneutron star into regions of lower temperature
and density, it first experiences an $\alpha$-particle
freeze-out, in which essentially all the protons are
consumed, followed by an $\alpha$-process (Woosley
\& Hoffman\markcite{woosley2} 1992), in which seed
nuclei near $A\sim 90$ are produced. The $r$-process
then takes place through the capture of the excess
neutrons on these seed nuclei.

During the dynamic phase of the $r$-process, a set of
progenitor nuclei are populated along the $r$-process
path through neutron capture, photodisintegration,
and charge-changing reactions. 
In the presence of
an intense neutrino flux as in the hot bubble,
the charge-changing
reactions include $\nu_e$ capture in addition to
the usual $\beta$-decay.
Due to the high temperature and the high neutron number
density at the hot bubble $r$-process site, the neutron
capture and photodisintegration reactions occur much
faster than the charge-changing weak reactions.
Consequently, within a given isotopic chain of charge $Z$, 
the relative abundances of 
the progenitor nuclei on the $r$-process
path are determined by the statistical 
$(n,\gamma)\rightleftharpoons(\gamma,n)$ equilibrium
(see e.g., Kratz et al.\markcite{kratz} 1993).
The relative progenitor abundances 
corresponding to isotopic chains 
at different $Z$ are governed by the
charge-changing weak reactions. When the neutron number
density drops below a critical level, the rapid neutron
capture stops and the progenitor abundance pattern
freezes out. 
The final $r$-process abundance
distribution is subsequently reached through a series
of charge-changing weak reactions that typically
conserve the nuclear mass number $A$. However, 
$\beta$-delayed and neutrino-induced neutron emission
changes $A$ and must be included in the transformation
from the neutron-rich progenitor nuclei to the
observed stable $r$-process nuclei.

\subsection{A simplified $r$-process calculation}
Various extensive $r$-process network calculations exist
in the literature (see e.g., Meyer et al.\markcite{meyer1}
1992; Kratz et al.\markcite{kratz} 1993). However, the
underlying key physics in such network calculations can
be elucidated with much more modest efforts. In this paper
we adopt the following simplified $r$-process calculation.
We start with only neutrons and seed nuclei, and further
assume that all seed nuclei have charge $Z_s=34$ and
mass number $A_s=90$ typically found for the products
of the $\alpha$-process (Hoffman et al.\markcite{hoffman}
1997). We then choose an $r$-process path.
Under the assumption of 
$(n,\gamma)\rightleftharpoons(\gamma,n)$ equilibrium,
the $r$-process path approximately follows the contour
of a constant neutron binding energy specified by the
temperature and the neutron number density 
(Kratz et al.\markcite{kratz} 1993). In general, this
path shifts during the $r$-process as both the temperature
and the neutron number density decrease with time. Rather than
relying on the assumption of
$(n,\gamma)\rightleftharpoons(\gamma,n)$ equilibrium
and keeping track of the change in the $r$-process path,
we choose an average nucleus with mass number $A_Z$ to
represent the progenitor nuclei in the isotopic chain
of charge $Z$. In fact, the typical $r$-process path,
especially the part at the magic neutron numbers,
does not rely on the particular assumption of
$(n,\gamma)\rightleftharpoons(\gamma,n)$ equilibrium.
We note that for a relatively low neutron 
number density of $\sim 10^{20}$--$10^{21}$ cm$^{-3}$,
the $r$-process path goes through a number of common
progenitor nuclei at the magic neutron numbers even if
the temperature is not high enough to establish an
$(n,\gamma)\rightleftharpoons(\gamma,n)$ equilibrium
(Cameron et al.\markcite{cam2} 1983). Thus, for simplicity,
we assume in this paper that there is a fixed $r$-process 
path with a unique relation between the progenitor charge
$Z$ and the corresponding mass number $A_Z$. 
It will become clear later that
this relation is used only when we evaluate the
neutron-to-seed ratio corresponding to a specific
abundance pattern for the progenitor nuclei at freeze-out.

At the magic
neutron number $N=82$, the average nuclei 
on the $r$-process path have charges
$Z=45$--49, corresponding to $A_Z=127$--131. Those at
the magic neutron number $N=126$ have charges $Z=65$--69,
corresponding to $A_Z=191$--195. We use a simple linear
interpolation to give $A_Z$ for the average nuclei
with non-magic neutron numbers
at $Z=35$--44 and 50--64. Because the solar $r$-process
abundances at $A>209$ (e.g., the actinides)
are very small, we assume that
all the abundances for $A>195$ are concentrated in an
average nucleus with $A=202$ as explained later. 
The chosen $r$-process path is shown in Fig. 1.
(The progenitor nuclei with magic neutron number
$N=50$ are not included in our simplified calculation,
because our assumed seed nuclei 
have neutron number $N_s>50$. In the
hot bubble $r$-process model the $N=50$ progenitor
nuclei are produced
in the $\alpha$-process. Consequently, the solar
$r$-process abundance peak at $A\sim 80$ usually
attributed to these progenitor nuclei will not
be discussed in this paper, which focuses on
the $r$-process nuclei near and above $A\sim 130$.)

Finally, we 
specify the $\beta$-decay rates for these average nuclei
on the $r$-process path. For the nuclei at the
$N=82$ and 126 closed neutron shells, we take the
$\beta$-decay rates from Table 4 in 
Fuller \& Meyer\markcite{fuller} (1995) and the tabulation
by M\"oller et al.\markcite{moller} (1996). 
The average $\beta$-decay rates are $\sim 4$ and 16 s$^{-1}$
for the progenitor nuclei with $N=82$ and 126, 
respectively. Although
several nuclei near $N=82$ and $A=130$ have experimentally
measured $\beta$-decay half-lives, the $\beta$-decay
properties for the majority of the progenitor nuclei
have to be calculated by theory, and therefore are subject
to considerable uncertainties. For our simplified 
$r$-process calculation, we take an approximate 
$\beta$-decay rate $\lambda_\beta\approx 50$ s$^{-1}$
for all the average nuclei with non-magic neutron numbers
(i.e., those with $Z=34$--44 and 50--64).
This rate is reasonable for the progenitor nuclei with
non-magic neutron numbers on a typical $r$-process path
whether $(n,\gamma)\rightleftharpoons(\gamma,n)$ equilibrium
is assumed or not.
In fact, our conclusions do not depend
sensitively on the particular choice of this rate as long as
it is much larger than the $\beta$-decay rates for the
progenitor nuclei with magic neutron numbers.

As mentioned previously, the intense neutrino flux in
the hot bubble necessitates the inclusion of
$\nu_e$ capture as an important type of charge-changing
reactions during the supernova $r$-process. 
Furthermore, we must include
$\nu_e$ capture in our $r$-process calculation in order to
consistently study various effects of this intense neutrino
flux on the $r$-process. The $\nu_e$ capture rates in an expanding
mass element depend on the $\nu_e$ flux, and hence on the
$\nu_e$ luminosity $L_{\nu_e}$ and the radius $r$ of the
mass element. Assuming that the $\nu_e$ luminosity evolves
with time $t$ as $L_{\nu_e}(t)=L_{\nu_e}(0)\exp(-t/\tau_\nu)$,
and that the mass element expands with a constant dynamic
timescale $\tau_{\rm dyn}$, i.e., 
$r(t)=r(0)\exp(t/\tau_{\rm dyn})$, we can write the rate for
$\nu_e$ capture on an average nucleus with charge $Z$ as
\begin{equation}
\lambda_{\nu_e}(Z,t)=\lambda_0(Z){L_{\nu_e,51}(0)\over
r_7(0)^2}\exp\left(-{t\over\hat\tau}\right),
\end{equation}
where $L_{\nu_e,51}$ and $r_7$ stand for $L_{\nu_e}$ in
unit of $10^{51}$ erg s$^{-1}$ and $r$ in unit of $10^7$ cm,
respectively, and
\begin{equation}
\hat\tau\equiv {\tau_{\rm dyn}\over 2}
{1\over 1+\tau_{\rm dyn}/(2\tau_\nu)}.
\end{equation}
In equation (1), $t=0$ is the time at which the $r$-process begins
in the mass element expanding away from the protoneutron star,
and $\lambda_0(Z)$ is the $\nu_e$ capture rate
for $L_{\nu_e}=10^{51}$ erg s$^{-1}$ and $r=10^7$ cm. We follow
the calculations of Qian et al.\markcite{qian2} (1997), and take
$\lambda_0(Z)\approx5.5$, 5.7, 7.0, and 8.5 s$^{-1}$ for
$Z=34$--44, 45--49, 50--64, and 65--69, respectively.

From our previous discussion of this supernova $r$-process model, 
it follows that the abundances of
progenitor nuclei on the $r$-process path are determined by
\begin{equation}
\dot Y(Z_s,t)=-[\lambda_\beta(Z_s)+\lambda_{\nu_e}(Z_s,t)]Y(Z_s,t),
\end{equation}
and
\begin{equation}
\dot Y(Z,t)=[\lambda_\beta(Z-1)+\lambda_{\nu_e}(Z-1,t)]Y(Z-1,t)
-[\lambda_\beta(Z)+\lambda_{\nu_e}(Z,t)]Y(Z,t)
\end{equation}
for $Z>Z_s$. Clearly, the total abundance of all progenitor
nuclei satisfy
\begin{equation}
\sum_{Z\geq Z_s}Y(Z,t)=Y(Z_s,0),
\end{equation}
where $Y(Z_s,0)$ is the total number of seed nuclei at
the beginning of the $r$-process.
We can define an average mass number $\bar A(t)$ for these
progenitor nuclei through
\begin{equation}
\bar A(t)Y(Z_s,0)=\sum_{Z\geq Z_s}A_ZY(Z,t)
=A_sY(Z_s,t)+\sum_{Z>Z_s}A_ZY(Z,t).
\end{equation}
From mass conservation, we have
\begin{equation}
Y_n(t)+\bar A(t)Y(Z_s,0)=Y_n(0)+A_sY(Z_s,0),
\end{equation}
where $Y_n$ is the neutron abundance. When $Y_n(t)/Y(Z_s,0)$
becomes negligible at $t=t_{\rm FO}$, the rapid neutron capture
stops, and the progenitor abundance pattern freezes out.
The condition for freeze-out then reads
\begin{equation}
\bar A(t_{\rm FO})=
\sum_{Z\geq Z_s}A_Z{Y(Z,t_{\rm FO})\over Y(Z_s,0)}
\approx A_s +{n\over s},
\end{equation}
where $n/s=Y_n(0)/Y(Z_s,0)$ is the neutron-to-seed ratio.

From a set of parameters $L_{\nu_e,51}(0)/r_7(0)^2$,
$\hat\tau$, and $n/s$,
our simplified $r$-process calculation described above
can be carried out in a straightforward manner. 
The progenitor abundance pattern at freeze-out is obtained
by integrating equations (3) and (4) until equation (8) is 
satisfied. However, the motivation of this paper is to
explore the diversity of supernova $r$-process. Therefore,
instead of adopting parameters from some specific supernova
model, we treat $L_{\nu_e,51}(0)/r_7(0)^2$, $\hat\tau$, and
$n/s$ as free parameters. Our goal is then
to find the parameters that can lead to the specific
freeze-out progenitor abundance patterns discussed
in the next subsection.

\subsection{Constraints on the $r$-process calculation} 
By employing extensive network calculations in their
$r$-process studies, previous workers 
have obtained detailed freeze-out abundance
patterns for the progenitor nuclei and followed the subsequent
$\beta$-decay to stability after freeze-out. Thus they
can compare their final $r$-process abundance
distributions with the observed solar $r$-process
abundance data on a nucleus-by-nucleus basis in order
to derive the varying physical conditions (e.g., neutron
number density, temperature, and $r$-process timescale)
at the $r$-process site(s) (Kratz et al.\markcite{kratz} 
1993) or to demonstrate the virtues of an astrophysical
model for the $r$-process (Woosley et al. 1994).
With our simplified $r$-process calculation, we are
not able to make such a detailed comparison. 
Instead, we try to
relate the essential features of the observed solar
$r$-process abundance distribution to the freeze-out
progenitor abundance patterns in our calculation.

First of all, we only consider the $r$-process nuclei
with $A\geq 127$, and divide the solar $r$-process
abundance distribution into four regions: (I) the
$A\sim 130$ peak ($A=127$--130), (II) $A=131$--190,
(III) the $A\sim 195$ peak ($A=191$--195), and
(IV) $A>195$ (cf. Fig. 1). Using the solar $r$-process abundance
data deduced by K\"appeler,
Beer, \& Wisshak\markcite{kapp} (1989), we find that
the sum of abundances in each of the first three
regions satisfies
\begin{equation}
N_I:N_{II}:N_{III}\approx 3:3:1.
\end{equation}
The sum of abundances for $A=196$--209 is slightly
less than that for region III. Allowing for the
depletion of the actinides ($A>209$) through fission, 
we assume
\begin{equation}
N_{IV}\sim N_{III}.
\end{equation}
In general, the solar $r$-process abundances result from
a superposition of different kinds of $r$-process events.
Since the sum of abundances in each region is not
affected very much by either $\beta$-delayed or
neutrino-induced neutron emission, we take, for example,
$N_I\propto\sum_ix_iY_I^i$, where $x_i$ is a weighting
parameter, and $Y_I^i$ is
the sum of the progenitor abundances in region I
at freeze-out in the $i$th kind of $r$-process events.
As only the
sum of abundances in region IV is of interest,
we just need to calculate $Y(Z,t)$ for
$Z_s\leq Z\leq 69$ in each kind of $r$-process events, 
and then obtain 
$\sum_{Z>69}Y(Z,t)$, and hence $N_{IV}$,
from equation (5).
Using the solar $r$-process abundance data,
we find that the average mass number for region IV is
about 202. The constraints in equations (9) and (10)
apply to any $r$-process scenario that yields the
observed solar $r$-process abundance pattern.

The next constraint, which distinguishes our calculation
from all earlier treatments,
takes into account the meteoritic
data on $^{129}$I/$^{127}$I and $^{182}$Hf/$^{180}$Hf. 
As stated in the
introduction, Wasserburg et al.\markcite{wass} (1996)
argued that the last $r$-process event contributing to
the $^{182}$Hf in the early solar system could make
only very little $^{129}$I.
Their argument applies to both the case where the
$r$-process nucleosynthesis was uniform over the
Galactic history and the case where the $^{129}$I and
$^{182}$Hf in the early solar system came only from 
the last supernova contribution to the protosolar system 
within $\sim 10^7$ yr of its formation.
Wasserburg et al.\markcite{wass} (1996) showed that
the amount of $^{182}$Hf in the early solar system
is consistent with a uniform production scenario,
which is also good for the actinides. According to
this scenario, the last $r$-process event responsible
for the $^{129}$I in the early solar system should have
occurred long ($\sim 10^8$ yr) before the last injection of
$^{182}$Hf, which took place within $\sim 10^7$ yr of
the solar system formation.
Consequently, there must be different $r$-process
sources for $^{129}$I and $^{182}$Hf. This difference
is possibly related to
a distinction between the 
$N=82$ and 126 closed neutron shells on the $r$-process
path.

Based on the argument of Wasserburg et al.\markcite{wass}
(1996), we consider the following minimal scenario.
We assume that there are two kinds of $r$-process events
contributing to the solar $r$-process abundances near
and above $A\sim 130$.
The first kind of events (case H) are
mainly responsible for the $r$-process nuclei near and above
$A\sim 195$ (regions III and IV).
They also make a significant amount of the
nuclei between $A\sim 130$
and 195 (region II), including $^{182}$Hf,
but very little $^{129}$I. The $r$-process nuclei
near $A\sim 130$ (region I) and the bulk
of those between $A\sim 130$ and 195
are made in the second kind of events (case L). 
In this scenario equations (9) and (10) can be rewritten as
\begin{equation}
(Y_I^H+xY_I^L):(Y_{II}^H+xY_{II}^L):(Y_{III}^H+xY_{III}^L):
(Y_{IV}^H+xY_{IV}^L)\approx 3:3:1:1,
\end{equation}
where for example, $Y_I^H$ is the sum of progenitor abundances
(normalized according to eq. [5])
in region I in case H, and $x$ is a weighting parameter
to be determined by our calculation. 
Physically, the weighting
parameter $x$ depends on the amount of $r$-process material
produced in a single event and the frequency of such events
in both cases H and L. 
Note that the quantities on the left-hand side of equation (11),
e.g., $Y_I^H+xY_I^L$, are proportional to
the sums of solar $r$-process abundances in the corresponding
regions, e.g., $N_I$, in equations (9) and (10). 

Ideally, we would like to have no production of $^{129}$I
at all in case H. Practically, we can set an upper limit on
the $^{129}$I production in case H as follows. We assume that
all the $^{129}$I in the early solar system was
produced by the $r$-process events in case H. This could be
realized if the period between the last $r$-process event
in case L and the solar system formation was long 
($\sim 10^8$ yr) compared
with the lifetime of $^{129}$I 
($\tau_{129}\approx 2.3\times 10^7$ yr).
We assume that this is
the case in the following discussion. Meteoritic 
measurements give the abundance ratio 
$^{129}{\rm I}/{^{127}{\rm I}} \approx 10^{-4}$ 
in the early solar system,
which corresponds to 
$^{129}{\rm I}/{^{195}{\rm Pt}}\approx 1.9\times 10^{-4}$.
In the uniform production scenario the abundance ratio
$^{129}{\rm I}/{^{195}{\rm Pt}}$ in the early solar system
is
\begin{equation}
{^{129}{\rm I}\over{^{195}{\rm Pt}}}\approx
{Y^H(Z=47,t_{\rm FO})\over Y^H(Z=69,t_{\rm FO})}{\tau_{129}\over t_G},
\end{equation}
where $Y^H(Z,t_{\rm FO})$ stands for $Y(Z,t_{\rm FO})$ in case H, and
$t_G\approx 10^{10}$ yr is the period of Galactic $r$-process
nucleosynthesis prior to the solar system formation.
The upper limit on the production of $^{129}$I in case H 
is then
\begin{equation}
{Y^H(Z=47,t_{\rm FO})\over Y^H(Z=69,t_{\rm FO})}\sim 0.1.
\end{equation}
In deriving the above upper limit we have assumed that the
final abundances of $^{129}$I and $^{195}$Pt are approximately
the same as the progenitor abundances for $(Z,A)=(47,129)$
and $(69,195)$ in case H. This assumption is reasonable because
the $\beta$-delayed neutron emission probabilities of the 
progenitor nuclei at and immediately above $A=129$ and 195
are small and neutrino-induced neutron emission after freeze-out
is severely constrained as discussed below.

Furthermore, it is believed that the abundance peaks
at $A\sim 130$ and 195 owe their existence to the
slow $\beta$-decay rates of the progenitor nuclei
at the $N=82$ and 126 closed neutron shells. In fact,
Kratz et al.\markcite{kra} (1988) showed that the
product of the freeze-out progenitor abundance at the closed
neutron shells and the corresponding
$\beta$-decay rate is approximately constant, i.e.,
a steady-state $\beta$-flow equilibrium approximately
holds for these progenitor nuclei at freeze-out.
Accordingly, we adopt the constraint
\begin{equation}
{\lambda_\beta(Z)Y(Z,t_{\rm FO})\over\lambda_\beta(Z+1)Y(Z+1,t_{\rm FO})}
\approx 1\pm 0.2
\end{equation}
for $Z=45$--47 (case L) and 65--68 (case H) in
our $r$-process calculation. As pointed out by Fuller \&
Meyer\markcite{fuller} (1995), the constraint in equation
(14) is especially important when $\nu_e$ capture is
included in the $r$-process calculation. It requires that
$\beta$-decay be the dominant charge-changing reaction
when the abundance peaks freeze out, i.e., it restricts
the $\nu_e$ flux at $t=t_{\rm FO}$.

Finally, we consider the effects of neutrino-induced neutron
emission after freeze-out. Qian et al.\markcite{qian2} (1997)
and Haxton et al.\markcite{haxton} (1997) showed that
neutrino-induced neutron emission results in significant
production of the nuclei in the valleys immediately below
the abundance peaks even for moderate neutrino fluences
after freeze-out. In order to produce the right amount 
of these nuclei,
the neutrino fluence ${\cal{F}}$
after freeze-out has to be sufficiently
low. For case H we have
\begin{equation}
{\cal{F}}={L_{\nu_e,51}(0)\over r_7(0)^2}
\exp\left(-{t_{\rm FO}\over\hat\tau}\right)\hat\tau\approx 0.015,
\end{equation}
and for case L
\begin{equation}
{\cal{F}}={L_{\nu_e,51}(0)\over r_7(0)^2}
\exp\left(-{t_{\rm FO}\over\hat\tau}\right)\hat\tau\approx 0.031.
\end{equation}
(The upper limits on ${\cal{F}}$
are 0.030 in case H and 0.045 in case L in order
not to overproduce these nuclei in the valleys.)

The constraints in equations (9)--(11) are 
treated in more accurate forms in earlier $r$-process
network calculations, but those in equations (13), (15), and (16)
have not been considered. While equation (14) is found to hold
in earlier $r$-process calculations (see e.g.,
Kratz et al.\markcite{kratz} 1993), its validity is essentially
guaranteed by the constraints in equations
(15) and (16) in our calculation.
In the future full network calculations 
will have to
be carried out in order to include
the $^{129}$I and $^{182}$Hf data and
allow for various neutrino effects. In this regard, our
simplified $r$-process calculation serves as an illustration
of the spirit, and hopefully, also as a stimulus for more
sophisticated future studies. 

\section{Results and Discussion}
As stated earlier,
there are three parameters $L_{\nu_e,51}(0)/r_7(0)^2$,
$\hat\tau$, and $n/s$ in our simplified $r$-process
calculation. Before we present the results of our
calculation, it is helpful to discuss the physics
that relates the set of these three parameters to the
progenitor abundance pattern at freeze-out in case H or L.
Obviously, in both
cases the neutron-to-seed ratio $n/s$ is related to
the average progenitor mass number $\bar A(t_{\rm FO})$
at freeze-out through equation (8). The mass number $A_Z$
of a progenitor nucleus is approximately proportional
to its charge $Z$, i.e., $A_Z\approx kZ$, where the
proportionality constant is
$k\approx 2.6$--2.9 for the $r$-process
path shown in Fig. 1. Therefore, the average progenitor charge
$\bar Z(t_{\rm FO})$ at freeze-out is
\begin{equation}
\bar Z(t_{\rm FO})=\sum_{Z\geq Z_s}Z
{Y(Z,t_{\rm FO})\over Y(Z_s,0)}\approx
{\bar A(t_{\rm FO})\over k}.
\end{equation}
From equations (8) and (17) we obtain
\begin{equation}
{n\over s}\approx k\bar Z(t_{\rm FO})-A_s,
\end{equation}
and we assume $k\approx 2.7$ in the following discussion.

Because only charge-changing reactions are involved
in equations (3) and (4), we can approximately view
the $r$-process as a charge flow proceeding from $Z_s$
to successively higher $Z$, accompanied by the capture
of $A_Z-A_s$ neutrons at each $Z$. When the neutrons
run out at $t=t_{\rm FO}$, the charges in the flow
have an average value $\bar Z(t_{\rm FO})$.
Without solving equations (3) and (4) for the charge flow,
we can approximately calculate this average progenitor charge 
$\bar Z(t_{\rm FO})$ at freeze-out as
\begin{equation}
\bar Z(t_{\rm FO})=Z_s+
\bar\lambda_\beta t_{\rm FO}+
\bar\lambda_{\nu_e}(0)\hat\tau\left[1-\exp
\left(-{t_{\rm FO}\over\hat\tau}\right)\right],
\end{equation}
where $\bar\lambda_\beta$ is the average $\beta$-decay
rate, and $\bar\lambda_{\nu_e}(0)$ is the average
initial $\nu_e$ capture rate [proportional to
$L_{\nu_e,51}(0)/r_7(0)^2$], both appropriately taken
for the progenitor nuclei involved in the calculation.
Equation (19) then relates 
$L_{\nu_e,51}(0)/r_7(0)^2$, $\hat\tau$, and the freeze-out
time $t_{\rm FO}$ to the progenitor abundance pattern
at freeze-out in both cases H and L. 

Furthermore,
$L_{\nu_e,51}(0)/r_7(0)^2$, $\hat\tau$,
and $t_{\rm FO}$ are subject to the neutrino fluence
constraints in equations (15) and (16) for cases H and L,
respectively. Therefore, one can
use either $L_{\nu_e,51}(0)/r_7(0)^2$ or $\hat\tau$ as
the only adjustable parameter in the two cases. Once
chosen, the other parameter, together with the freeze-out
time $t_{\rm FO}$, is determined by the 
average progenitor charge at freeze-out (eq. [19]) 
and the neutrino
fluence constraint (eq. [15] or [16]) in each case.

\subsection{Results for a given $L_{\nu_e,51}(0)/r_7(0)^2$}
For the convenience of presentation, we
first give results for a reasonable value of
$L_{\nu_e,51}(0)/r_7(0)^2\approx 8.77$, which corresponds
to $\lambda_{\nu_e}(Z,0)\approx 50$ s$^{-1}$ for the progenitor
nuclei with $N=82$. The dependence of our results on
$L_{\nu_e,51}(0)/r_7(0)^2$ will be examined in \S3.3.
Our best fit for case H is obtained for 
$\hat\tau\approx 0.186$ s and $n/s\approx 92$.
The corresponding freeze-out time is $t_{\rm FO}\approx 0.86$ s.
The time evolution of the progenitor abundance pattern in
case H is shown in Fig. 2 as a series of snapshots.
Similarly, the best fit for case L
is obtained for $\hat\tau\approx 0.125$ s and $n/s\approx 48$,
with the corresponding
freeze-out time $t_{\rm FO}\approx 0.44$ s. The time evolution
of the progenitor abundance pattern in case L is
shown in Fig. 3.
For a given $L_{\nu_e,51}(0)/r_7(0)^2$, we find that
case H is specified essentially by the constraint on
$^{129}$I production
in equation (13) and the neutrino fluence constraint
in equation (15). With the freeze-out pattern
obtained in case H, case L is specified by 
the solar $r$-process abundance ratios
in equation (11) and the neutrino fluence constraint
in equation (16). The weighting parameter
in equation (11) is found to be $x\approx 2.17$.
With the above best-fit parameters, all the constraints
discussed in \S2.2 are satisfied.

The abundance pattern obtained from the
superposition of cases H and L is shown in Fig. 4.
As explained previously, we cannot compare this pattern
with the solar $r$-process abundance distribution
on a nucleus-by-nucleus basis, especially because
we do not follow the transformation from the progenitor
nuclei to the stable $r$-process nuclei after freeze-out. 
However, if we assume that the progenitor nuclei
at the $N=82$ and 126 closed neutron shells approximately
conserve their mass numbers during the transformation
after freeze-out, the final abundances at these mass
numbers (regions I and III) shown in Fig. 4
agree with the solar 
$r$-process abundances in the $A\sim 130$ and 195 peaks
quite well. While we cannot obtain detailed abundances
for the $r$-process nuclei in regions II and IV
mainly due to significant $\beta$-delayed neutron emission
after freeze-out expected in these two regions, at least
the sums of the abundances in these two regions,
together with those in regions I and III, agree with
the solar $r$-process abundance pattern as required by
our calculation.

Furthermore, we can show that the abundance ratio
$^{182}$Hf/$^{180}$Hf in the early solar system is also
consistent with the meteoritic data and with the scenario
where the $r$-process events in both cases H and L occurred
uniformly up until the solar system formation. As explained
previously, the constraint on $^{129}$I 
production requires that the last
$r$-process event in case L contributing to the solar
abundances occur $\sim 10^8$ yr before the solar system
formation. Because the lifetime of $^{182}$Hf is
$\tau_{182}\approx 1.3\times 10^7\ {\rm yr}\ll 10^8$ yr,
the $^{182}$Hf made in this last $r$-process event in case L
had already decayed to the stable $^{182}$W when the
solar system was formed. Following 
Wasserburg et al.\markcite{wass} (1996), we take
$^{182}{\rm W}_r/{^{180}{\rm Hf}}\approx 0.37$, where
$^{182}{\rm W}_r$ represents the $r$-process contribution
to the solar abundance of $^{182}$W. In the uniform production
scenario we have 
\begin{equation}
{^{182}{\rm Hf}\over^{182}{\rm W}_r}\approx
{Y_{182}^H\over Y_{182}^H+xY_{182}^L}{\tau_{182}\over t_G},
\end{equation}
where $Y_{182}^H$ and $Y_{182}^L$ stand 
for the final abundances
of $^{182}$Hf in cases H and L, respectively.
Assuming $Y_{182}^H\sim Y_{182}^L$ (cf. Figs. 2 and 3), we 
obtain $^{182}{\rm Hf}/^{182}{\rm W}_r\sim4.1\times 10^{-4}$,
which corresponds to 
$^{182}{\rm Hf}/^{180}{\rm Hf}\sim 1.5\times 10^{-4}$ in good
agreement with the meteoritic value
$^{182}{\rm Hf}/^{180}{\rm Hf}\approx 2.8\times 10^{-4}$.
In fact, we can always obtain this agreement as long as
$Y_{182}^H\gtrsim Y_{182}^L$ and $x\sim 1$.

We now examine the effect of $\nu_e$ capture on the
charge flow. In our calculation 
equations (15) and (16), which concern the
neutrino fluence after freeze-out,
impose much more stringent constraints on the $\nu_e$ flux
than equation (14), which concerns the approximate $\beta$-flow
equilibrium at freeze-out. This result was found earlier
by Qian et al.\markcite{qian2}
(1997). By the time the progenitor abundance pattern
freezes out, the charge flow is carried dominantly by
$\beta$-decay in both cases H and L.
However, whereas equation (14) is satisfied for all
five progenitor nuclei ($Z=65$--69) in the $N=126$ peak
in case H, it is satisfied only for three progenitor
nuclei ($Z=46$--48) in the $N=82$ peak in case L. This is because
the bottle-neck in the charge flow due to the slow
$\beta$-decay rates for the $N=82$ progenitor
nuclei facilitates the establishment of an
approximate $\beta$-flow equilibrium in the $N=126$ peak
in case H,
whereas no corresponding bottle-neck exists before the
$N=82$ progenitor nuclei in case L. On the other hand,
$\nu_e$ capture accelerates the charge flow quite noticeably
in both cases H and L. 

We recall that for given values of 
$L_{\nu_e,51}(0)/r_7(0)^2$ and $\hat\tau$,
equation (19) determines the freeze-out time
$t_{\rm FO}$ as a function of the average progenitor charge
$\bar Z(t_{\rm FO})$ at freeze-out.
Here we give a more accurate way to evaluate
this function. The time $\delta t(Z)$ required for the charge
flow to pass through the progenitor nucleus at charge $Z$ is
approximately determined by
\begin{equation}
\lambda_\beta(Z_s)\delta t(Z_s)+\lambda_{\nu_e}(Z_s,0)\hat\tau
\left\{1-\exp\left[-{\delta t(Z_s)\over\hat\tau}\right]\right\}
\approx 1,
\end{equation}
and
\begin{equation}
\lambda_\beta(Z)\delta t(Z)+\lambda_{\nu_e}[Z,t(Z)]\hat\tau 
\left\{1-\exp\left[-{\delta t(Z)\over\hat\tau}\right]\right\}
\approx 1
\end{equation}
for $Z>Z_s$,
where $t(Z)=\sum_{Z'=Z_s}^{Z-1}\delta t(Z')$ is the time required 
for the charge flow to proceed
from $Z_s$ up to $Z$. We assume that the freeze-out time
$t_{\rm FO}$ is approximately given by
\begin{equation}
t_{\rm FO}=t[\bar Z(t_{\rm FO})]
\approx\sum_{Z=Z_s}^{\bar Z(t_{\rm FO})-1}\delta t(Z),
\end{equation}
with $t_{\rm FO}=0$ for $\bar Z(t_{\rm FO})=Z_s$.
It is easy to see that equation (19) is obtained by replacing
$\lambda_\beta(Z)$ and $\lambda_{\nu_e}(Z,0)$ for $Z\geq Z_s$
with $\bar\lambda_\beta$ and $\bar\lambda_{\nu_e}(0)$ in 
equations (21) and (22). Using equations (21)--(23),
we plot $t_{\rm FO}$ as a function of the average
progenitor charge $\bar Z(t_{\rm FO})$
at freeze-out for 
$L_{\nu_e,51}(0)/r_7(0)^2\approx 8.77$, $\hat\tau\approx 0.186$
and 0.125 s, and for the case without neutrinos in Fig. 5.
The time required to reach the same average progenitor charge 
at freeze-out is clearly
longer without neutrinos than with neutrinos. The actual
freeze-out times $t_{\rm FO}$ in cases H and L are indicated
as filled circles in Fig. 5. The shortening of $t_{\rm FO}$ in
both cases with respect to the case without neutrinos 
(see \S3.2) mainly
results from the $\nu_e$-capture-induced
acceleration of the charge flow at
$Z=45$--49, i.e., at the progenitor nuclei with the $N=82$ closed
neutron shell.

To conclude this subsection,
we give a semi-analytic way to derive $\hat\tau$ and $n/s$
in cases H and L for a given $L_{\nu_e,51}(0)/r_7(0)^2$.
As discussed in the beginning of \S3, the neutron-to-seed
ratio $n/s$ is approximately given by the average
progenitor charge $\bar Z(t_{\rm FO})$ at freeze-out via
equation (18). From the solar $r$-process abundance ratios in
equation (11) and the constraint on $^{129}$I production
in equation (13),
we see that a
large fraction of the progenitor abundances should be in
region III (I) at freeze-out in case H (L).
Consequently,
the average progenitor charge at freeze-out has to be
$\bar Z(t_{\rm FO})\approx 68$--69 (48--49) in case H (L),
which requires a neutron-to-seed ratio of $n/s\approx 94$--96
(40--42) in good agreement with our numerical results.
Once $\bar Z(t_{\rm FO})$ is known, the freeze-out time
$t_{\rm FO}$ can be calculated as a function of $\hat\tau$ for
a given $L_{\nu_e,51}(0)/r_7(0)^2$ using equations (21)--(23). 
The contours for
$\bar Z(t_{\rm FO})=48$, 49, 68, and 69 are 
shown as solid lines
on the $\hat\tau$-$t_{\rm FO}$ plot in Fig. 6. 
Furthermore, the neutrino fluence constraint in equation (15)
or (16) gives $t_{\rm FO}$ as another function of $\hat\tau$
for a given $L_{\nu_e,51}(0)/r_7(0)^2$. The contours
for values of the neutrino fluence after freeze-out
${\cal{F}}=0.015$ and 0.031 are shown as dashed lines
in Fig. 6. The best-fit parameters for $t_{\rm FO}$ and 
$\hat\tau$ in case H (L) should then
lie on the dashed line for ${\cal{F}}=0.015$
(0.031) and between the solid lines for 
$\bar Z(t_{\rm FO})=68$ (48) and
69 (49). This is confirmed by our numerical results, which
are indicated as filled circles in Fig. 6.

\subsection{Results for the case without neutrinos}
We now consider the case without neutrinos, i.e.,
$L_{\nu_e,51}(0)/r_7(0)^2=0$. Obviously, the neutrino
fluence constraints in equations (15) and (16)
can no longer be satisfied, and can only be treated
as some upper limits on the neutrino fluence after
freeze-out in this case. By leaving out neutrinos
and the associated constraints in equations (15) and (16),
we also find freeze-out progenitor abundance patterns
that can satisfy essentially all the other constraints
discussed in \S2.2. These freeze-out progenitor abundance
patterns corresponding to cases H$'$ and L$'$ are similar
to those in cases H and L presented in \S3.1, but are
obtained with slightly smaller neutron-to-seed ratios
and considerably longer freeze-out times. 
The neutron-to-seed ratio in case H$'$ (L$'$) is
$n/s\approx 86$ (44) with the corresponding freeze-out
time $t_{\rm FO}\approx 1.68$ (0.78) s. As shown in
Fig. 5, approximately the same average progenitor charge
at freeze-out is reached in cases H and H$'$ or
in cases L and L$'$.
The weighting parameter in equation (11) is $x\approx 1.11$ 
for case L$'$ with respect to case H$'$ in
order to give the best fit to the gross solar $r$-process abundance
pattern. The time evolution of
the progenitor abundance pattern in the case without neutrinos is
shown in Fig. 7.

Here we notice some interesting differences between
cases H$'$ (L$'$) and H (L). Case H$'$
is essentially determined by the constraint on $^{129}$I
production in equation (13). Due to
the slow $\beta$-decay rates for the progenitor nuclei with $N=82$,
a long $t_{\rm FO}$ is required to decrease the progenitor abundance
at $A=129$. However, once the charge flow passes the bottle-neck
at $N=82$, it reaches the progenitor nuclei at $A>195$ relatively 
fast. Consequently, the $r$-process nuclei at $A>195$ are overproduced
by about 40\% in case H$'$ in order to satisfy the
constraint on $^{129}$I production in equation (13). 
By comparison, the decaying $\nu_e$
flux in case H has the beneficial effect of accelerating the 
passage through the bottle-neck at $N=82$ at an earlier time without
overproducing the $r$-process nuclei at $A>195$ at later times.
Furthermore, without neutrinos
the approximate $\beta$-flow equilibrium constraint
in equation (14) is satisfied only for
two progenitor nuclei $(Z=47$--48) in the $N=82$ peak in 
case L$'$. Therefore,
while we cannot conclude that neutrinos are required to satisfy
all the constraints derived from the observed solar $r$-process 
abundance data,
the cases with neutrinos seem to be
more attractive.

\subsection{Dependence on $L_{\nu_e,51}(0)/r_7(0)^2$}
We have presented the results for a fixed value of
$L_{\nu_e,51}(0)/r_7(0)^2\approx 8.77$ in \S3.1 and
for the case without neutrinos corresponding to
$L_{\nu_e,51}(0)/r_7(0)^2=0$ in \S3.2.
We now examine the dependence of our results on
$L_{\nu_e,51}(0)/r_7(0)^2$ while taking into account
all the constraints discussed in \S2.2. In other words,
we want to find those cases that are similar to case
H or L, but have different values of $L_{\nu_e,51}(0)/r_7(0)^2$.

As explained at the end of \S3.1, the solar $r$-process
abundance ratios in equation (11) and the constraint on
$^{129}$I production in equation (13) require that the
average progenitor charge at freeze-out be
$\bar Z(t_{\rm FO})=68$--69 (48--49) in case H (L).
Consequently, the neutron-to-seed ratio $n/s$ in those
cases similar to case H (L) has to be close to
94--96 (40--42). The other two parameters in our calculation,
$L_{\nu_e,51}(0)/r_7(0)^2$ and $\hat\tau$, together with
the freeze-out time $t_{\rm FO}$, are constrained by
the average progenitor charge $\bar Z(t_{\rm FO})$ at
freeze-out and the neutrino fluence ${\cal{F}}$ after
freeze-out (eq. [15] or [16]) in each case. Therefore,
the combination of $L_{\nu_e,51}(0)/r_7(0)^2$ and $\hat\tau$
in those cases similar to case H would most likely be located
in the region between the contour lines for
$[\bar Z(t_{\rm FO}),{\cal{F}}]=(68,0.015)$ and $(69,0.015)$
on the $\hat\tau$ vs. $L_{\nu_e,51}(0)/r_7(0)^2$ plot.
Likewise, the combination of $L_{\nu_e,51}(0)/r_7(0)^2$ and $\hat\tau$
in those cases similar to case L would most likely be located
in the region between the contour lines for
$[\bar Z(t_{\rm FO}),{\cal{F}}]=(48,0.031)$ and $(49,0.031)$
on the same plot. This plot is shown as Fig. 8. Obviously,
the parameter regions shown in Fig. 8 include the best-fit
parameters in cases H and L.
We have checked a number of other combinations of
$L_{\nu_e,51}(0)/r_7(0)^2$ and $\hat\tau$ within these regions, 
and have confirmed that
they give similar results to those discussed previously.
In particular, the results corresponding to
$L_{\nu_e,51}(0)/r_7(0)^2\sim 1$ are very close to
those in the case without neutrinos.

Furthermore, although the parameter $\hat\tau$ in case L
is shorter than that in case H for the same 
$L_{\nu_e,51}(0)/r_7(0)^2$, we can find a case L$''$ that
has a smaller $L_{\nu_e,51}(0)/r_7(0)^2$ and a longer
$\hat\tau$ than both cases H and L (see Fig. 8), and at the
same time, gives a freeze-out progenitor abundance
pattern essentially identical to that in case L. Specifically,
the parameters are $L_{\nu_e,51}(0)/r_7(0)^2\approx 1.75$
[corresponding to $\lambda_{\nu_e}(Z,0)\approx 10$ s$^{-1}$
for the progenitor nuclei with $N=82$], 
$\hat\tau\approx 0.25$ s, and $n/s\approx 47$ in case L$''$.
The corresponding freeze-out time is $t_{\rm FO}\approx 0.66$ s.
With the same weighting parameter $x\approx 2.17$ in equation
(11), cases H and L$''$ give the same 
best-fit to the gross solar
$r$-process abundance pattern as cases H and L.
It follows that a range of $L_{\nu_e,51}(0)/r_7(0)^2$ and
$\hat\tau$ within the regions shown in Fig. 8 can provide
the yields in cases H and L.

\section{Conclusions}
We have found that the gross solar $r$-process abundance
pattern near and above $A\sim 130$
can be reproduced by a superposition of two kinds
of supernova $r$-process events after taking into account
the meteoritic data
on $^{129}$I and $^{182}$Hf.
The first kind of events (case H)
are mainly responsible for the $r$-process nuclei near and
above $A\sim 195$. They also make a significant amount
of the nuclei between $A\sim 130$ and 195, including
$^{182}$Hf, but very little $^{129}$I. The $r$-process
nuclei near $A\sim 130$ and the bulk of those between
$A\sim 130$ and 195 are made in the second kind of
events (case L). In each case, the $r$-process
nucleosynthesis in a mass element expanding away from
the protoneutron star is governed by the initial $\nu_e$ flux
$L_{\nu_e,51}(0)/r_7(0)^2$ at the beginning of the $r$-process,
the decay timescale $\hat\tau$ of the $\nu_e$ flux, and the
neutron-to-seed ratio $n/s$.
The parameter $n/s$ specifies the
$r$-process nuclei mainly produced in each case. The other
two parameters, $L_{\nu_e,51}(0)/r_7(0)^2$ and $\hat\tau$, are
important in determining when all the neutrons are used up,
i.e., the freeze-out time $t_{\rm FO}$. Therefore, they
determine the neutrino fluence ${\cal{F}}$ after freeze-out,
which may be responsible for the production of certain $r$-process
nuclei through neutrino-induced neutron emission 
(Qian et al.\markcite{qian2} 1997; 
Haxton et al.\markcite{haxton} 1997). In addition,
the $\nu_e$ flux 
plays a significant, possibly even crucial role in decreasing
the production of $^{129}$I with respect to $^{182}$Hf
in case H. In both cases H and L, the solar $r$-process
abundance ratios in equation (11) and the constraint on
$^{129}$I production in equation (13) determine the average
progenitor charge $\bar Z(t_{\rm FO})$ at freeze-out, and hence
the neutron-to-seed ratio $n/s$. For a given
$L_{\nu_e,51}(0)/r_7(0)^2$, the parameter $\hat\tau$,
together with the freeze-out time $t_{\rm FO}$, can be
calculated from $\bar Z(t_{\rm FO})$ and the neutrino fluence
constraint (eq. [15] or [16]) for each case, 
as shown in Fig. 6. The dependence of $\hat\tau$ on
$L_{\nu_e,51}(0)/r_7(0)^2$ is shown in Fig. 8.

We wish to emphasize that the meteoritic
constraint on coproduction of $^{129}$I with
$^{182}$Hf leads to well-defined parameters, especially
the neutron-to-seed ratio, in case H. 
As illustrated by the case without
neutrinos, it is difficult to suppress the production
of $^{129}$I, which has an $N=82$ progenitor nucleus
with a long $\beta$-decay lifetime, and to avoid 
overproduction of the $r$-process nuclei at $A>195$
at the same time. When $\nu_e$ capture is included
in the $r$-process calculation, this difficulty is
noticeably alleviated. However, the parameters characterizing
the $\nu_e$ flux are then subject to additional constraints.
Consequently,
case H represents a particular kind of $r$-process events
with possibly a very narrow range of neutron-to-seed ratios
($n/s\sim 90$).
On the other hand, although we have shown that
the gross solar $r$-process abundance pattern near and
above $A\sim 130$ can be accounted for in
the minimal scenario of two kinds of $r$-process events,
the progenitor abundance pattern in case L can be regarded
as some average over different events spanning a broader range of
neutron-to-seed ratios (e.g., $n/s\sim 40$--50), 
as long as these events occur infrequently
enough to be consistent with the meteoritic data on
$^{129}$I and $^{182}$Hf. According to
Wasserburg et al.\markcite{wass} (1996), the events
in case H occur roughly once every $10^7$ yr, whereas
those represented by case L occur roughly once
every $10^8$ yr within a region of $\sim 100$ pc in size
in the Galaxy.

The size of $\sim 100$ pc may be understood from the
expansion of the supernova ejecta. For an explosion
energy of $\sim 10^{51}$ erg, the initial velocity
of the supernova ejecta is $v_0\sim 10^3$ km s$^{-1}$.
In about a few $10^3$ yr, the supernova sweeps over
a distance of about 6 pc and mixes with about the same
amount of the interstellar medium as the total mass
$M_{\rm ej}$ of the original ejecta. At times
$t\gg 10^3$ yr, the expansion (commonly
known as ``snow plowing'') under momentum conservation
is described by
\begin{eqnarray}
R&\approx&\left({3\over\pi}{M_{\rm ej}v_0t_{\rm exp}\over\rho_{\rm ISM}}
\right)^{1/4}\nonumber\\
&\approx& 63\left({t_{\rm exp}\over 10^7\ {\rm yr}}\right)^{1/4}
\left({M_{\rm ej}\over 20\ M_\odot}\right)^{1/4}
\left({v_0\over 2\times 10^3\ {\rm km\ s}^{-1}}\right)^{1/4}
\left({m_H\ {\rm cm}^{-3}\over\rho_{\rm ISM}}\right)^{1/4}\ {\rm pc},
\end{eqnarray}
where $R$ is the radius of the expansion front from the center
of the supernova, $t_{\rm exp}$ is the expansion time since the supernova
explosion, $\rho_{\rm ISM}$ is the density of the 
interstellar medium, and $m_H$ is the mass of the hydrogen atom.
If we ignore other means of mixing such as Galactic rotation,
clearly only those supernovae that could reach the position
of the protosun within the lifetime of $^{182}$Hf 
($\tau_{182}\approx 1.3\times 10^7$ yr) were responsible
for the $^{182}$Hf in the early solar system. From equation (24),
those supernovae occurred within $\sim 70$ pc from the position
of the protosun for reasonable values of the relevant parameters. 
Interestingly, within the lifetime of $^{182}$Hf,
the number of supernovae in a region of $\sim 70$ pc in size is
about one assuming a total Galactic volume of $\sim 700$ kpc$^3$
and a supernova frequency of $\sim (30\ {\rm yr})^{-1}$ 
in the whole Galaxy. Therefore, $^{182}$Hf
can be replenished on a timescale of $\sim 10^7$ yr
consistent with the meteoritic data if the supernova
$r$-process events in case H occur with a frequency of
$f_{\rm SN}^H\sim (30\ {\rm yr})^{-1}$ in the whole Galaxy,
as also argued by Wasserburg et al.\markcite{wass} (1996).

On the other hand, the meteoritic data require that the $^{129}$I
produced along with $^{127}$I be replenished on a much longer
timescale of $\sim 10^8$ yr. Because
the lifetimes of $^{129}$I and $^{182}$Hf are very close,
the regions enclosing the supernovae contributing to these
two nuclei have about the same size. Consequently, those
supernova $r$-process events represented by case L must occur
with a frequency of $f_{\rm SN}^L\sim (300\ {\rm yr})^{-1}$ 
in the whole Galaxy. The frequencies $f_{\rm SN}^H$ and
$f_{\rm SN}^L$, together with the corresponding amounts of
$r$-process ejecta $M_r^H$ and $M_r^L$ in cases H and L,
determine the superposition parameter $x$ in equation (11).
As found in \S3,
we have $x\sim (M_r^L/M_r^H)(f_{\rm SN}^L/f_{\rm SN}^H)\sim 1$--2.
So the amount of $r$-process
material ejected in the less frequent case L 
is $\sim 10$--20 times more than
that in the more frequent case H. 
This implies that the mass loss rate is
much higher,
or more likely, that the period for ejecting $r$-process material
is much longer in case L than in case H.

Following the preceding arguments for two distinct $r$-process
sources, we propose the following $r$-process 
scenario assuming that all of the Type II supernovae producing
$r$-process nuclei are of a generally similar nature. We suggest
that material with higher neutron-to-seed ratios is ejected
in the neutrino-driven winds at higher neutrino luminosities,
i.e., at earlier times
during the neutrino cooling phase of the protoneutron
star. In addition, the early $r$-process ejecta have a
neutron-to-seed ratio of $n/s\sim 90$. The neutron-to-seed ratio
then rapidly decreases to $\sim 40$--50.
If neutrino emission were uninterrupted, the neutron-to-seed ratio
would stay $\sim 40$--50, and the corresponding
amount of material, all ejected, would be
$\sim 10$--20 times more than the amount of material with 
$n/s\sim 90$. However, we consider that the continuous mass loss
in the neutrino-driven winds is commonly terminated 
during the rapid transition from $n/s\sim 90$ to $\sim 40$--50.
This would occur
in $\sim 90$\% of the Type II supernovae, with only $\sim 10$\%
of them having prolonged continuous mass loss. 
Depending on the initial core mass of the
supernova progenitor, among other things,
both neutrino emission and mass loss could be
terminated by black hole formation during the neutrino cooling
phase of the protoneutron star (Brown \& Bethe\markcite{brown} 1994).
In this scenario
there would then be $\sim 5\times 10^8$ black holes with masses
$\sim 1\ M_\odot$ from the $r$-process events in case H and
$\sim 5\times 10^7$ neutron stars from the less frequent 
$r$-process events represented by
case L in the Galaxy today.

In the above $r$-process scenario 
we have associated high neutron-to-seed ratios
with high neutrino luminosities and low neutron-to-seed ratios
with low neutrino luminosities. Qualitatively, a shorter
$\hat\tau$ is expected for a higher neutrino luminosity
(Qian \& Woosley\markcite{qian1} 1996). 
This can be achieved in the framework of the present model
(cf. cases H and L$''$ in Fig. 8).
Of course, a consistent
set of the three parameters $L_{\nu_e,51}(0)/r_7(0)^2$, $\hat\tau$,
and $n/s$ at different times during the neutrino cooling phase
of the protoneutron star can only be obtained in a detailed
numerical study of Type II supernovae.

We note that many other nuclear species are
produced by the explosive nucleosynthesis (e.g., Fe and Si)
in Type II supernovae and by the hydrostatic burning 
(e.g., $^{16}$O) in the outer envelope during the presupernova
evolution.
The explosive nucleosynthesis is associated
with the shock propagation through the envelope.
The products from both the explosive nucleosynthesis and
the hydrostatic burning are largely
unaffected by the neutrinos from the protoneutron star
[except for the $\nu$-process discussed by 
Woosley et al.\markcite{woosley4} (1990)]
or by the possible formation of a black hole during the
neutrino cooling phase of the protoneutron star.
Therefore, the abundant non-$r$-process nuclei
are ejected together
with the $r$-process elements 
in a Type II supernova. The scenario given here would
not significantly alter the usual supernova contributions
to the non-$r$-process nuclei.

Furthermore, we note that neutrino-driven winds also develop
after the accretion-induced collapse (AIC) of a white
dwarf into a neutron star
(Woosley \& Baron\markcite{woosley1} 1992). Therefore,
the AIC events could also correspond to the infrequent $r$-process
events represented by case L. However, because there is no
envelope around the final neutron star in the AIC events,
the overall nucleosynthetic signature of such events is different
from that of Type II supernovae. Only the nuclear species
produced in the neutrino-driven winds, especially the
$r$-process nuclei, are ejected in the AIC events.

Finally, the diversity of $r$-process sources have
some interesting consequences for Galactic chemical
evolution. At very low metallicities, only Type II supernovae
could make Fe, whereas both Type Ia and Type II
supernovae contribute to Fe at sufficiently high metallicities.
Therefore,
if the $r$-process events in case H were mainly
associated with Type II supernovae,
the abundance ratio
of the corresponding main $r$-process product 
with respect to Fe would remain constant at low metallicities
and decrease with increasing metallicity after Type Ia
supernovae began to make Fe.
On the other hand, if the $r$-process events represented by
case L were mainly associated with the AIC events, the
metallicity dependence for the abundance ratio
of the corresponding main $r$-process product 
with respect to Fe would be 
sensitive to the difference between the time
at which
such events first occurred and the onset of increase in metallicity 
due to Type Ia supernova.

The above suggestions regarding the nature of supernova
$r$-process events are highly speculative.
However, if the binary distribution of $r$-process sources with
very different frequencies and very different mass contributions
is correct, then some new supernova
$r$-process models will be required along the general lines
indicated here. 

\acknowledgments
We want to thank Gerry Brown and Sterl Phinney for helpful
discussions. This work was supported in part by the U. S.
Department of Energy under Grant No. DE-FG03-88ER-40397,
by NASA under Grant No. NAG5-4076, and by Division
Contribution No. 5827(973).
Y.-Z. Qian was supported by the David W. Morrisroe 
Fellowship at Caltech.

\vfill
\eject
\figcaption{The average $r$-process path used in the simplified
calculation. The progenitor nuclei on the path are shown as open
squares on the charge ($Z$) vs. neutron number ($N$) plot. The
filled squares indicate the typical nuclei at $\beta$-stability.
The progenitor nuclei are divided into four groups, corresponding
to the four regions in the solar $r$-process nuclear abundance
distribution: $A=127$--130 (I), 131--190 (II), 191--195 (III),
and $A>195$ (IV). The average mass number in region IV is
$\langle A\rangle =202$. As a reminder, the magic neutron numbers
are shown explicitly.}

\figcaption{The time evolution of the progenitor abundance pattern
in case H, assuming $L_{\nu_e,51}/r_7(0)^2\approx 8.77$, 
$\hat\tau\approx 0.186$ s, and $n/s\approx 92$.
The sum of all progenitor abundances is normalized to unity. 
The progenitor abundances at $A=129$ and 191--195 are shown as
filled circles. The sum of progenitor abundances at $A>195$ is
indicated as the filled circle at $A=202$. The progenitor abundance
pattern freezes out at $t=t_{\rm FO}\approx 0.86$ s and
satisfies all the constraints discussed in \S2.2.}

\figcaption{The time evolution of the progenitor abundance pattern
in case L, assuming 
$L_{\nu_e,51}/r_7(0)^2\approx 8.77$, $\hat\tau\approx 0.125$ s,
and $n/s\approx 48$. The sum of all progenitor abundances is 
normalized to unity.
The progenitor abundances at $A=127$--130 and 191--195
are shown as filled circles.  
The sum of progenitor abundances at $A>195$ is
indicated as the filled circle at $A=202$.
The progenitor abundance pattern freezes out at 
$t=t_{\rm FO}\approx 0.44$ s and
satisfies all the constraints discussed in \S2.2.}

\figcaption{The abundance pattern that matches the bulk solar
$r$-process abundances in regions I, II, III, and IV 
for the minimal two-component model (see text). 
This pattern
is obtained by superposing the freeze-out
progenitor abundance pattern in case H
with that in case L. The freeze-out 
progenitor abundance pattern
in case L is weighted by a factor of 2.17 with respect to that in
case H. The abundances at $A=127$--130 and 191--195
are shown as filled circles.  
The sum of abundances at $A>195$,
indicated as the filled circle at $A=202$, is chosen to be
0.1. All the other abundances are scaled accordingly.}

\figcaption{The freeze-out time $t_{\rm FO}$ as a function of
the average progenitor charge $\bar Z(t_{\rm FO})$ at freeze-out
for $L_{\nu_e,51}/r_7(0)^2\approx 8.77$,
$\hat\tau\approx 0.186$ and 0.125 s, and
for the case without neutrinos.
The actual freeze-out time 
$t_{\rm FO}\approx 0.86$ (0.44) s in case
H (L) is indicated as the filled circle labelled H (L). 
Note that approximately
the same average progenitor charge at freeze-out as in case H (L) 
is reached at a considerably longer freeze-out time
$t_{\rm FO}\approx 1.68$ (0.78) s in the case without neutrinos 
[the open circle labelled H$'$ (L$'$)].}

\figcaption{The determination of $\hat\tau$ and $t_{\rm FO}$ in cases H 
and L for $L_{\nu_e,51}/r_7(0)^2\approx 8.77$. 
The solid lines give the contours for values of the average
progenitor charge at freeze-out
$\bar Z(t_{\rm FO})=48$, 49, 68, and 69, 
as calculated from equations (21)--23,
on the $\hat\tau$-$t_{\rm FO}$ plot.
The dashed lines give the contours for values of the neutrino fluence
after freeze-out ${\cal{F}}=0.015$ and 0.031,
as calculated from equations (15) and (16), on the 
same plot. The filled circles labelled H and L
indicate the best-fit parameters 
in cases H and L, respectively, which satisfy the corresponding equations
for $\bar Z(t_{\rm FO})$ and ${\cal{F}}$.}

\figcaption{Same as Fig. 3, but for the case without neutrinos.
Freeze-out progenitor abundance patterns similar to those
in cases H and L are
obtained with neutron-to-seed ratios $n/s\approx 86$ and 44
at $t=t_{\rm FO}\approx 1.68$ and 0.78 s, respectively.}

\figcaption{The parameter regions for $L_{\nu_e,51}(0)/r_7(0)^2$
and $\hat\tau$ in those cases similar to cases H and L. 
The solid lines correspond to the contours for the following
values of the average progenitor charge 
at freeze-out and the neutrino fluence after freeze-out:
$[\bar Z(t_{\rm FO}),{\cal{F}}]=(48,0.031)$, $(49,0.031)$, 
$(68,0.015)$ and $(69,0.015)$.
For a neutron-to-seed ratio $n/s\sim 94$--96 (40--42), 
the parameters $L_{\nu_e,51}(0)/r_7(0)^2$ and $\hat\tau$ in
those cases similar to case H (L) are most likely located
in the region between the solid lines that includes the filled
circle labelled H (L). For example, the combination of
$L_{\nu_e,51}(0)/r_7(0)^2$ and $\hat\tau$ indicated as the
filled circle labelled L$''$ is shown. This combination
gives essentially the same
freeze-out progenitor abundance pattern as in case L,
but with a lower $L_{\nu_e,51}(0)/r_7(0)^2$ and a longer $\hat\tau$.
Consequently, we can choose generic cases H and L lying in
the corresponding regions in this figure
to give the same best-fit to the gross solar
$r$-process abundance pattern. All the constraints discussed
in \S2.2 would be satisfied by these choices.}

\vfill
\eject
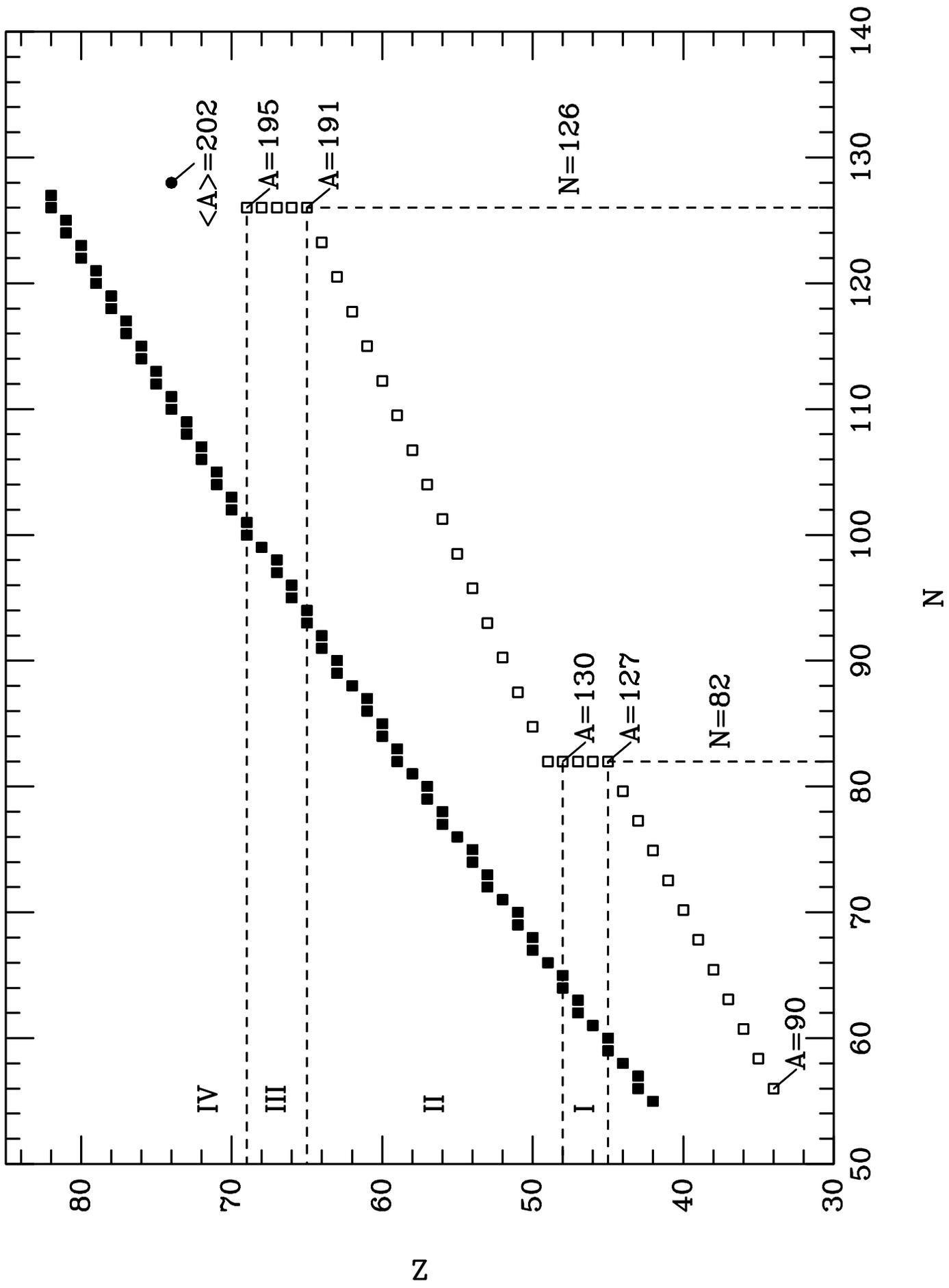
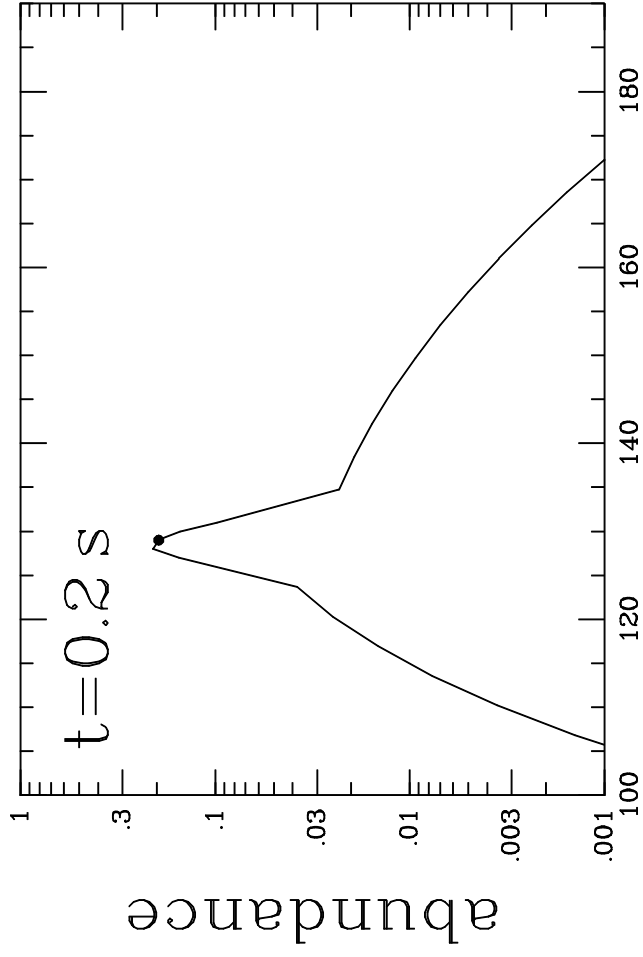
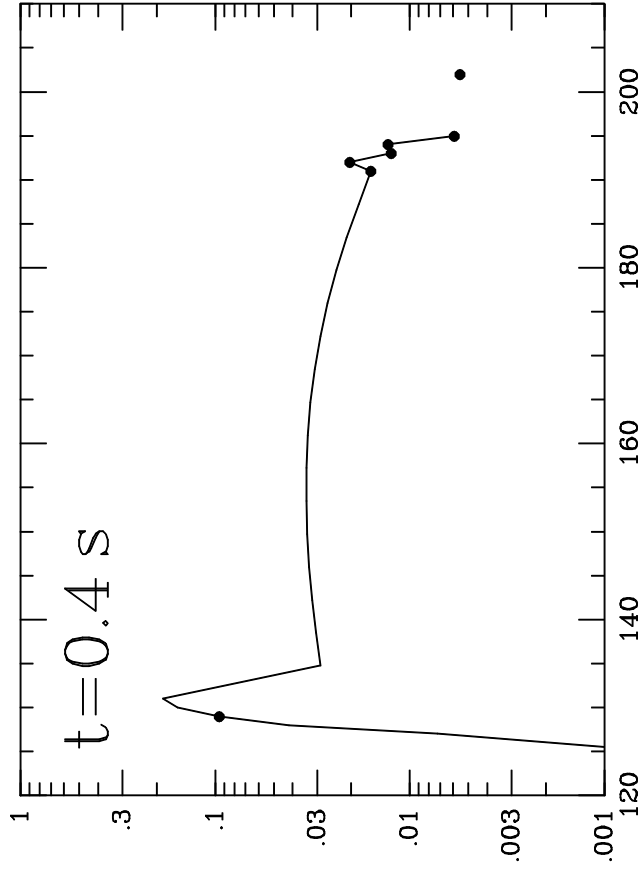
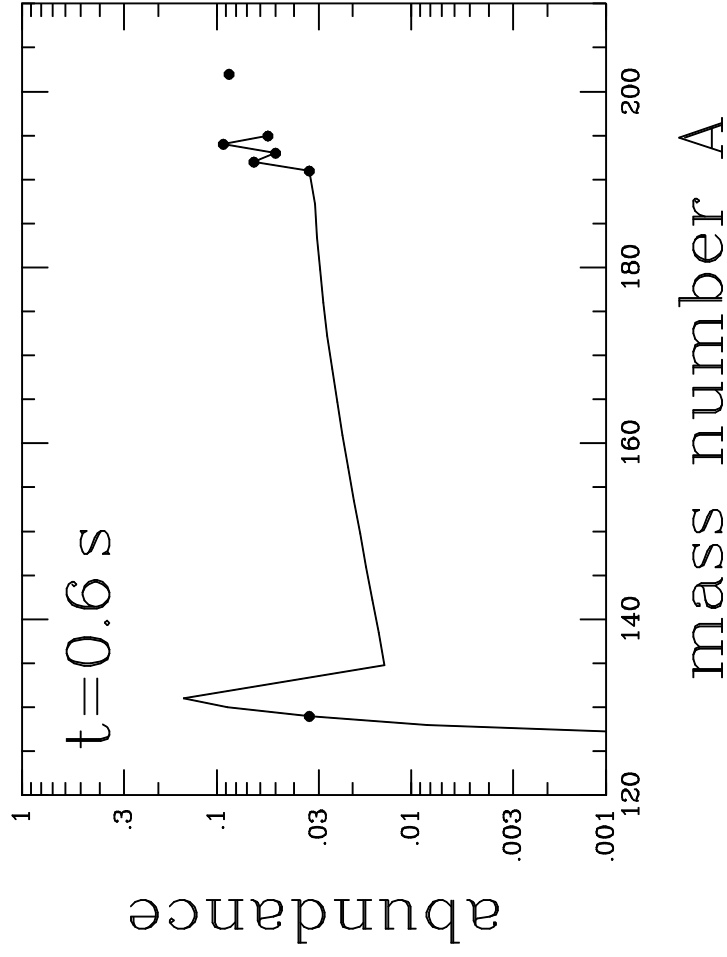
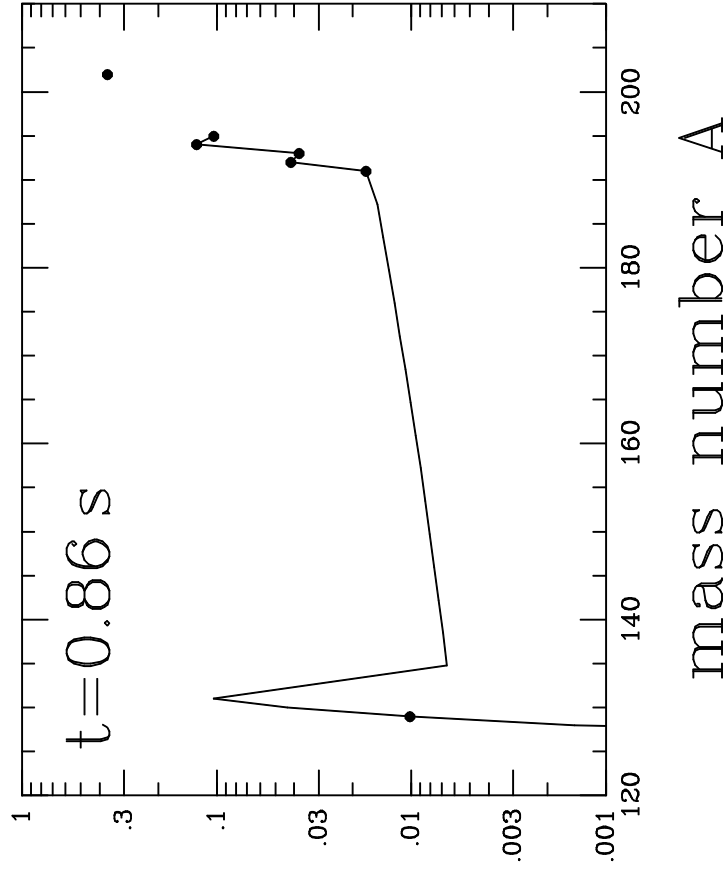
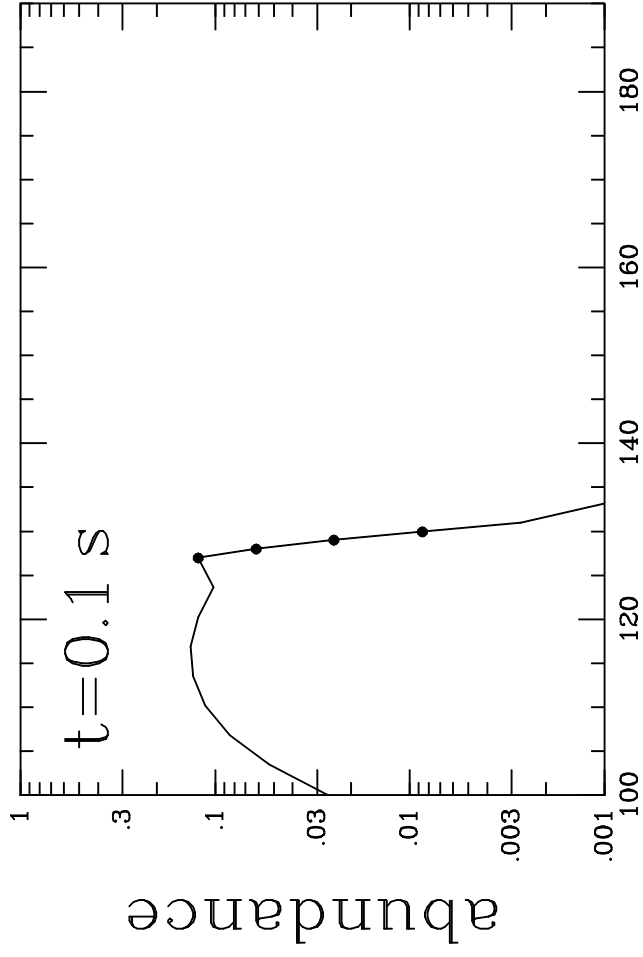
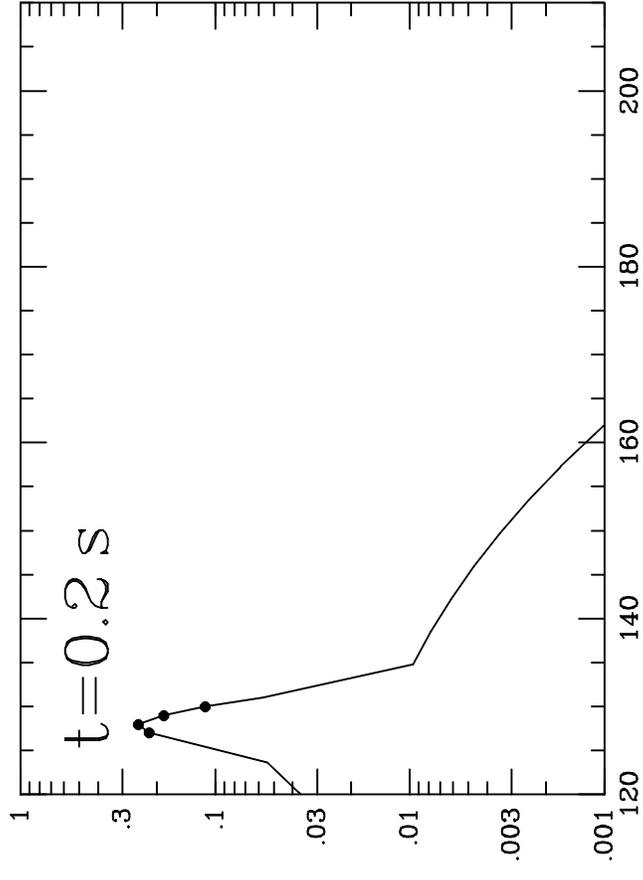
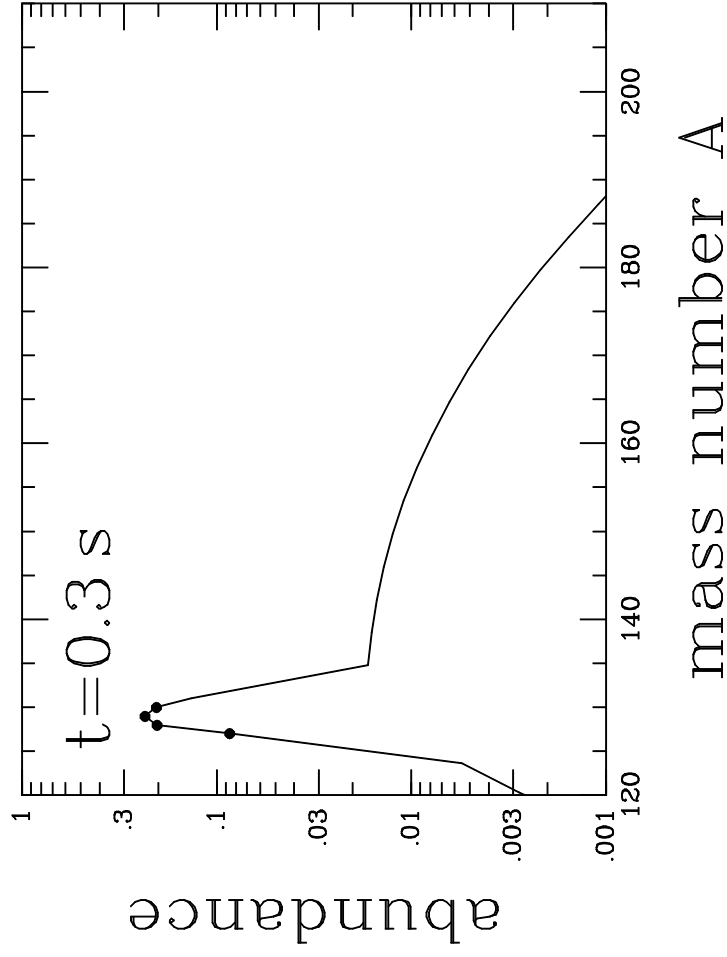
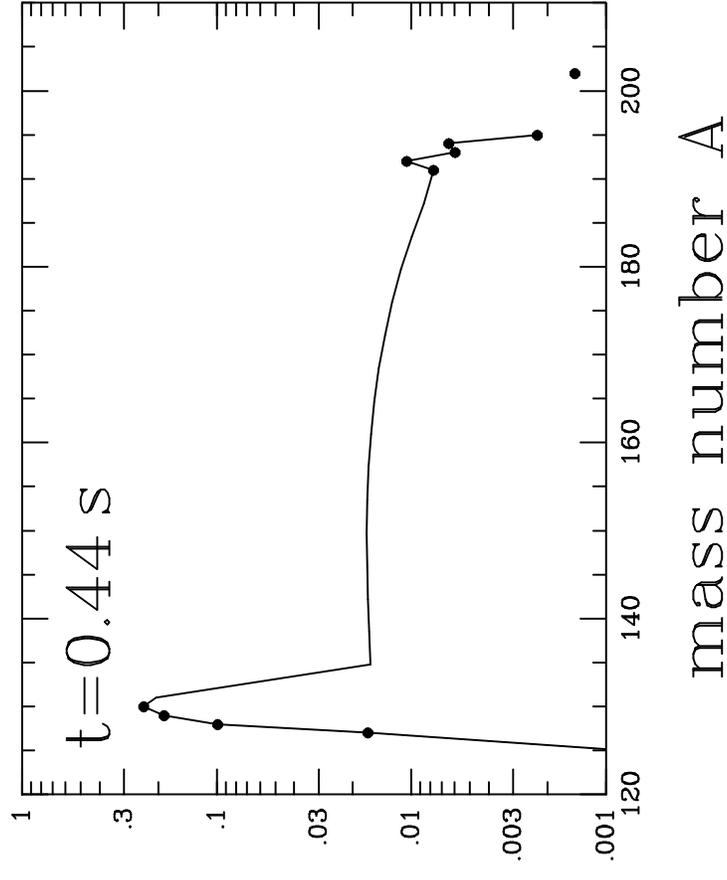
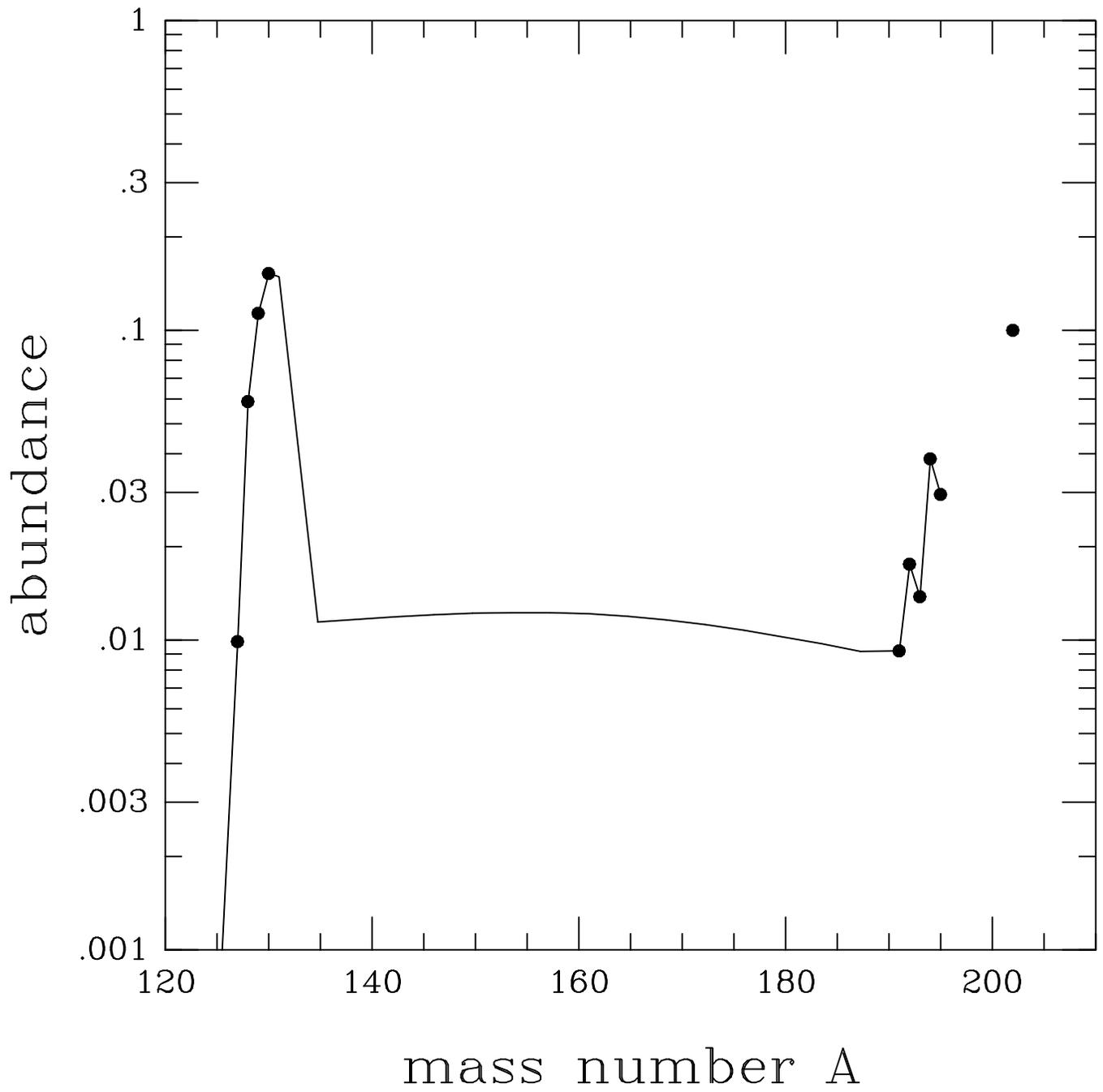
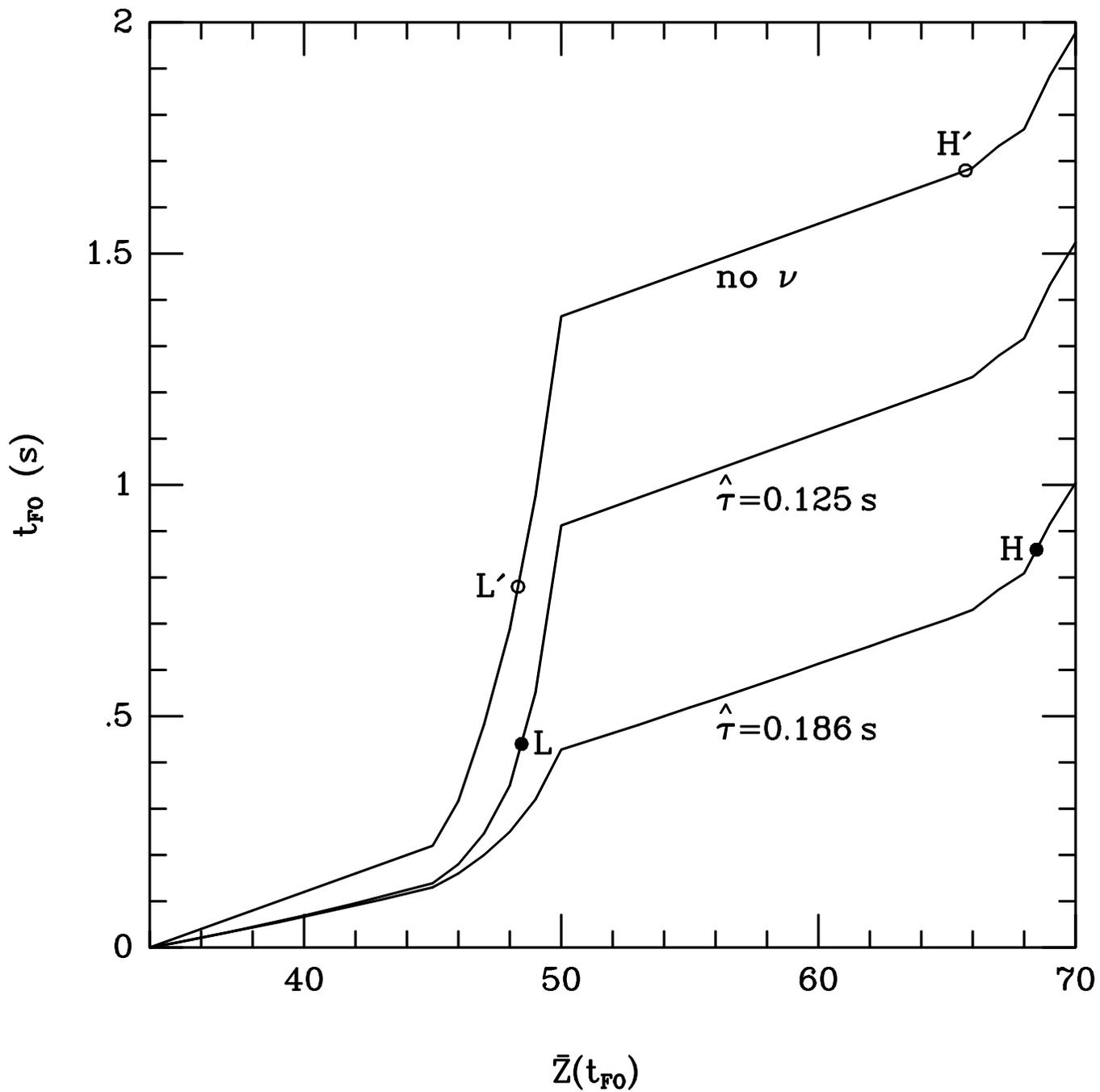
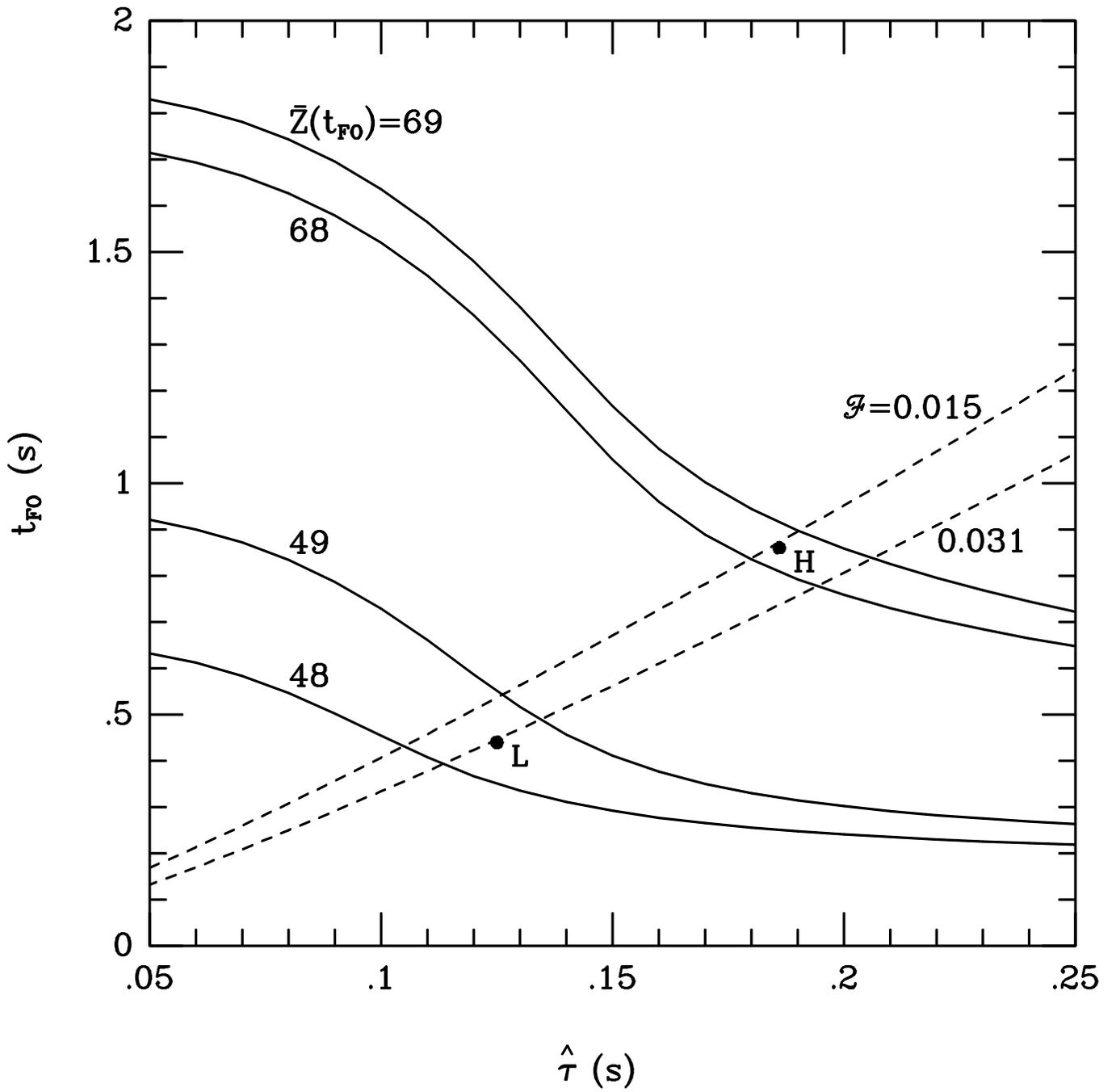
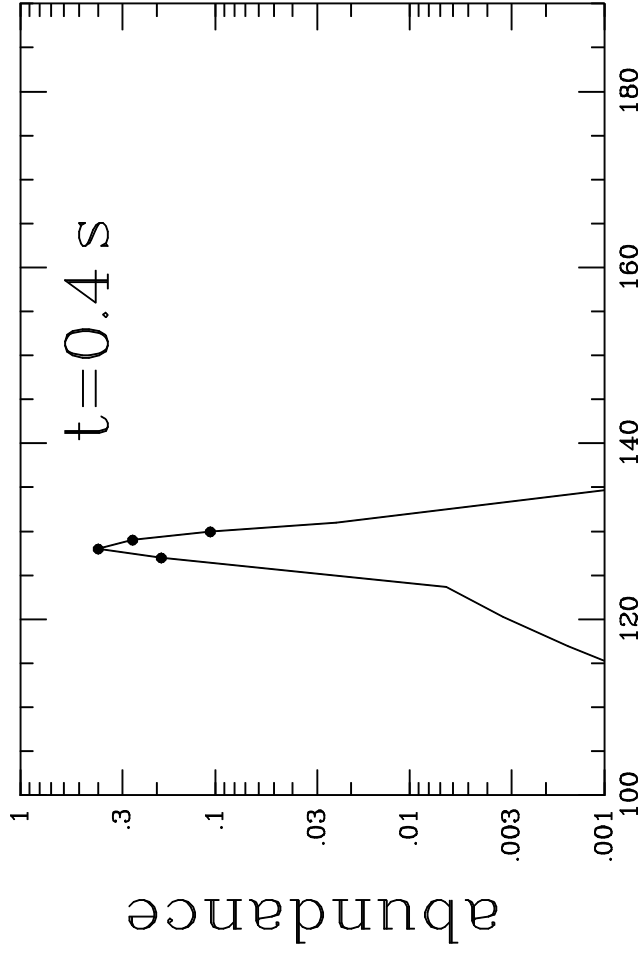
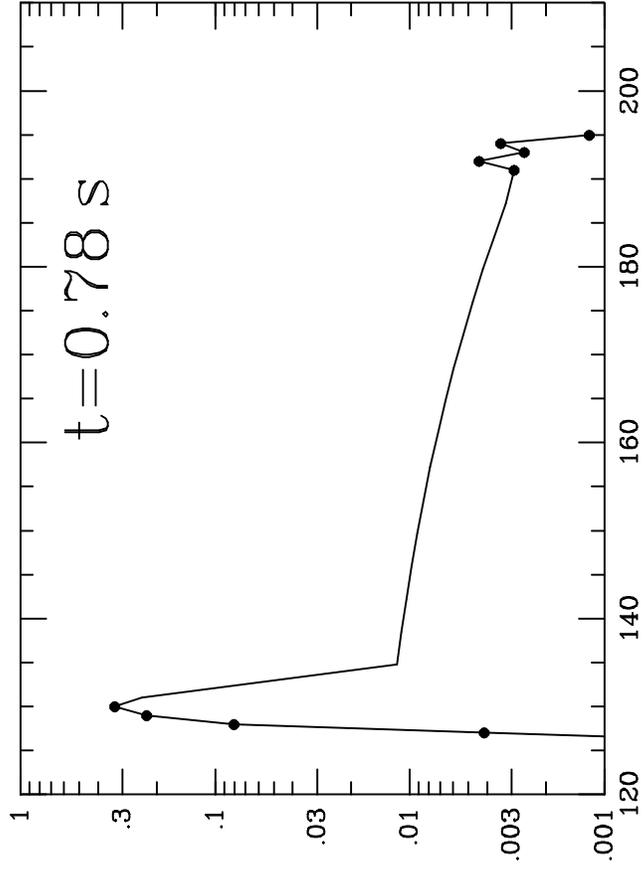
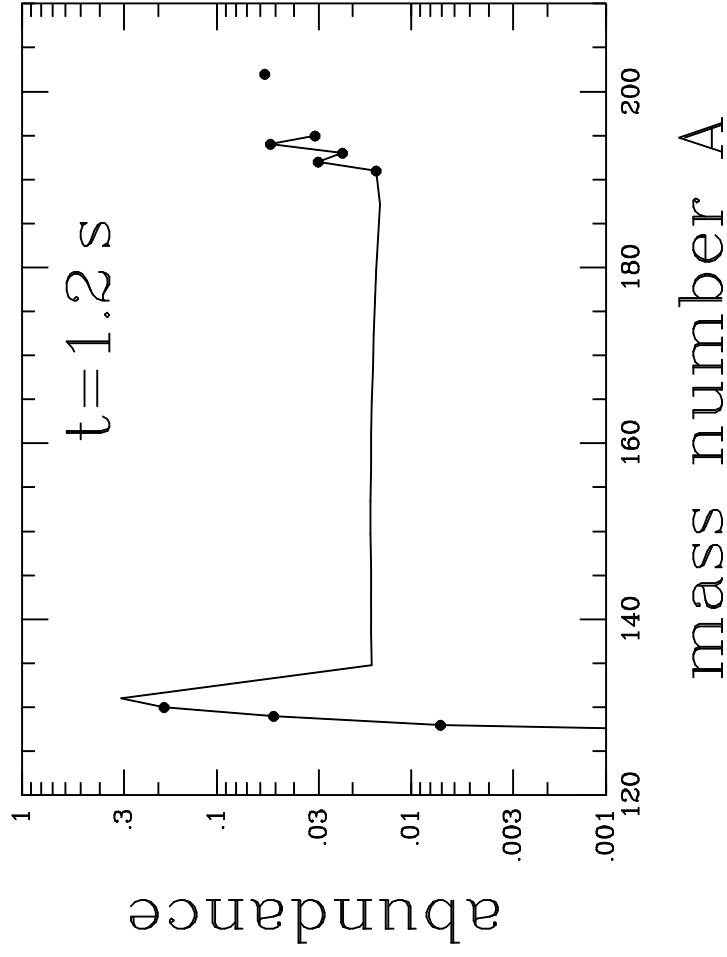
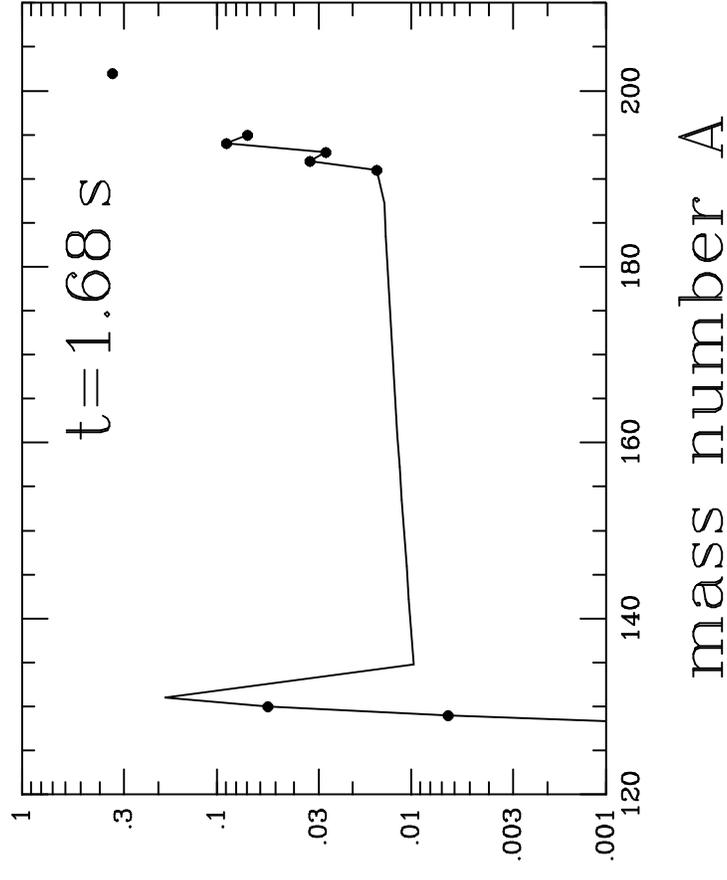
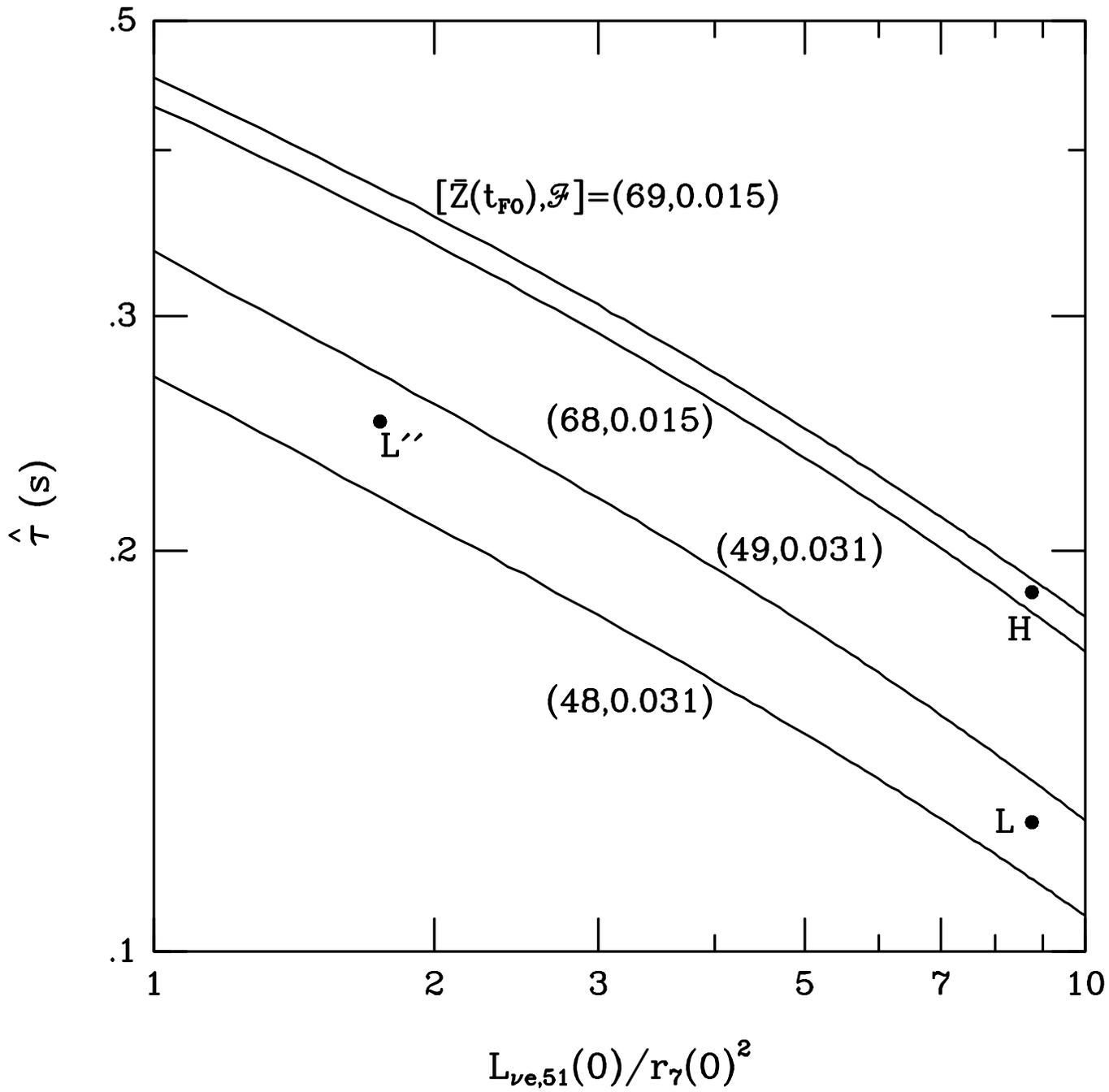
\end{document}